\documentclass[11pt, a4paper]{article}       
\usepackage[utf8]{inputenc}
\usepackage[british]{babel}
\usepackage[nottoc]{tocbibind}
\usepackage{authblk}

\usepackage{blindtext}
\usepackage[top=1in, bottom=1in,left=1in,right=1in]{geometry}
\usepackage{amssymb}
\usepackage{amsmath}
\usepackage{mathtools}
\usepackage{multicol}
\usepackage{tikz}

\usepackage{amsfonts}
\usepackage{graphicx}
\usepackage{subfig}
\usepackage{caption}
\usepackage{braket}
\usepackage{array}
\usepackage{float}
\usepackage{hyperref}
\usepackage{cite}
\newcolumntype{M}[1]{>{\centering\arraybackslash}m{#1}}
\newcolumntype{N}{@{}m{0pt}@{}}
\usepackage{enumerate}
\usepackage{bm}
\usepackage{bbm}
\usepackage{mathptmx}  
\usepackage{mathrsfs}
\providecommand{\keywords}[1]
{
  \small	
  \textbf{\textit{Keywords---}} #1
}

\begin{document}

\title{\textbf{Pure and Mixed State Entanglement Dynamics in Tavis–Cummings Model with Squeezed Coherent Thermal States}}

\def\correspondingauthor{\footnote{Corresponding author's email: mandalkoushik1993@gmail.com}}
\author[]{Koushik Mandal\correspondingauthor{}}

\author[]{M. V. Satyanarayana}

\affil[]{\textit{Department of Physics, Indian Institute of Technology Madras, Chennai, India, 600036}}

\date{}

\maketitle


\begin{abstract}
We investigate the entanglement dynamics of two atoms interacting with a single-mode cavity field within the Tavis--Cummings model in the presence of noise. The atoms are initially prepared in either pure Bell states or mixed Werner states, allowing a direct comparison of pure- and mixed-state entanglement. The cavity field is described by generalized single-mode squeezed coherent thermal states, incorporating both thermal and quantum noise effects.

Atom--atom and atom--field entanglement are quantified using concurrence and negativity, respectively. We analyze entanglement sudden death and revival, and examine how Ising-type coupling, dipole--dipole interaction, Kerr nonlinearity, and detuning modify the entanglement dynamics. Our results show that thermal photons generally suppress entanglement and enhance sudden death, while squeezing counteracts these effects. The influence of nonlinearities and interatomic interactions depends sensitively on the purity of the initial atomic state, leading to qualitatively different behaviors for Bell and Werner states.
\end{abstract}


\section{Introduction}

Quantum entanglement \cite{RevModPhys.81.865} is a central resource in quantum information theory \cite{isar1994open, PhysRevA.64.062106, PhysRevA.70.052110, PhysRevLett.99.160502, PhysRevA.83.022109, xu2013experimental, PhysRevB.90.054304, dajka2014disentanglement, Aolita_2015, PhysRevA.92.012315}, enabling fundamental tasks such as quantum teleportation \cite{PhysRevLett.70.1895}, superdense coding \cite{PhysRevLett.69.2881}, quantum secret sharing, and anonymous communication \cite{PhysRevLett.67.661}. Understanding the dynamics of entanglement in atom-field systems is crucial for both fundamental studies and practical implementations in quantum information processing. Over the years, significant efforts have been devoted to entanglement dynamics in systems such as trapped ions \cite{PhysRevA.76.041801, PhysRevA.95.043813, PhysRevA.97.043806, PhysRevA.49.1202}, cavity QED \cite{PhysRevA.82.053832, rosadon, AHMADI2011820, niemczyk2010circuit, haroche2020cavity, fink2008climbing, PhysRevLett.105.173601, PhysRevA.92.023810}, circuit QED \cite{Mamgain_2023}, and linear optical platforms. These systems naturally involve interactions between atoms and quantized electromagnetic fields, making the study of atom-field entanglement essential.

The Jaynes-Cummings model (JCM) \cite{jaynes1963comparison}, describing a two-level atom interacting with a single-mode field, remains one of the most fundamental models in quantum optics. It captures essential phenomena such as Rabi oscillations, collapse and revival, and nonclassical features of the field. Experimentally, JCM physics has been explored in Rydberg atoms in high-Q cavities and in trapped ions \cite{nayak1988quantum, gea1990collapse, karatsuba2009resummation, cirac1994quantum, jakubczyk2017quantum, vogel1989k, Chong_2020}. In our previous work \cite{mandal2023atomic}, we analyzed how thermal and squeezed photons affect atomic inversion and entanglement for a single two-level atom within the JCM framework.

Extending to two atoms, the double Jaynes-Cummings model (DJCM) \cite{eberly} describes two independent cavities, each containing a two-level atom. This model exhibits entanglement sudden death (ESD), where atom-atom entanglement temporarily vanishes before reviving \cite{Yönaç_2006, PhysRevLett.97.140403, eberly2, YU2006393, Yönaç_2007, yu2005evolution, yu2009sudden, eberly3}. Prior studies of DJCM have considered both squeezed coherent states and Glauber-Lachs states, investigating how thermal and squeezed photons influence atoms initially prepared in pure (Bell) or mixed (Werner) states. Recent work has further explored squeezed coherent thermal states (SCTS) in DJCM and intensity-dependent DJCM, focusing on Bell-state atoms.

In contrast to the double Jaynes--Cummings model, where each atom interacts with an independent cavity mode, the present work considers two atoms \emph{collectively coupled} to a single--mode cavity field. This configuration corresponds to the two--atom Tavis--Cummings model, in which the atoms interact with the field through a common quantized mode.

Several studies have addressed additional factors affecting entanglement dynamics, including atomic spontaneous decay, cavity damping, detuning, and intensity-dependent interactions \cite{li2020entanglement, qin2012entanglement}. Furthermore, interatomic interactions such as Ising-type and dipole-dipole couplings, as well as Kerr-nonlinearity, can substantially modify entanglement evolution \cite{Pandit_2018, PhysRevA.101.053805, laha2023dynamics, obada2018influence}. These interactions can alter ESD patterns, induce entanglement transfer between subsystems, and generate complex oscillatory behaviors. 

Motivated by these observations, in this work we investigate the entanglement dynamics of two atoms collectively coupled to a single--mode cavity field (two--atom Tavis--Cummings model) with squeezed coherent thermal states, considering both Bell and Werner initial atomic states. Specifically, we aim to:
\begin{itemize}
    \item Analyze and compare entanglement dynamics for atoms in pure versus mixed states under the influence of thermal and squeezed photons.
    \item Examine the roles of Ising-type and dipole-dipole interactions on atom-atom and atom-field entanglement.
    \item Explore the impact of detuning and Kerr-nonlinearity on entanglement evolution and ESD phenomena.
    \item Provide physical insights into how initial states and interactions govern entanglement generation, decay, and transfer between subsystems.
\end{itemize}

\begin{figure}
    \centering
    \includegraphics[width=0.5\linewidth]{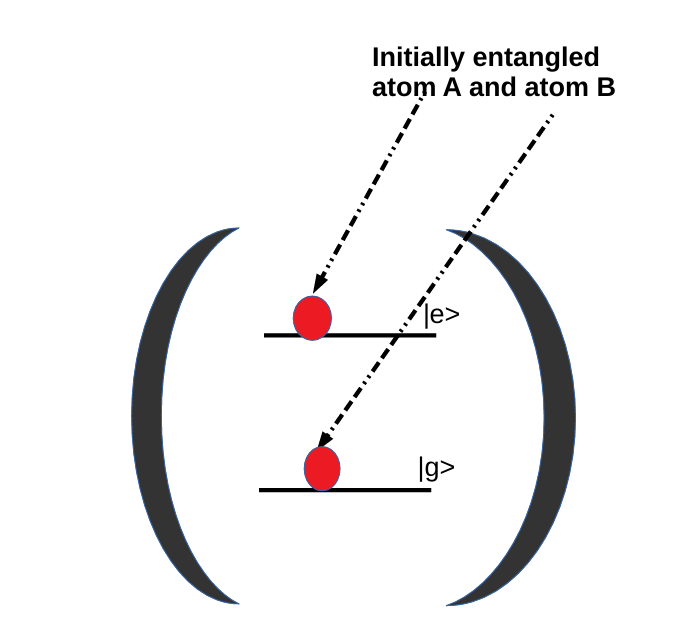}
    \caption{Schematic of two entangled atoms interacting with a single-mode cavity field.}
    \label{single_entang_jcm}
\end{figure}

A schematic of the system is shown in Fig.~\ref{single_entang_jcm}. The results presented here not only extend previous investigations to mixed atomic states and nonlinear effects but also offer insights relevant for quantum information applications in cavity QED and related platforms.

The organization of the paper is as follows: Section 2 introduces the radiation and atomic states, entanglement measures, and the theoretical model of atom-field interaction. Section 3 discusses the effects of thermal and squeezed photons. Sections 4 and 5 explore the impact of Ising-type and dipole-dipole interactions. Section 6 addresses detuning effects, Section 7 investigates Kerr-nonlinearity, and Section 8 summarizes the key findings and conclusions.

\section{Theory}
\subsection{Photonic state}
Squeezed coherent thermal states (SCTS) are mixed states because of the of thermal photons. The density operator for SCTS is defined as\cite{PhysRevA.47.4474, PhysRevA.47.4487, yi1997squeezed}

\begin{equation}
\hat{\rho}_{\textsubscript{SCT}} = \hat{D}(\alpha)\hat{S}(\zeta)\hat{\rho}_{\text{th}}\hat{S}^{\dagger}(\zeta)\hat{D}^{\dagger}(\alpha),
\label{rho_sct}
\end{equation}
where
\begin{equation}
\hat{D}(\alpha) = \exp(\alpha \hat{a}^{\dagger} - \alpha^{*} \hat{a})
\end{equation}
is the displacement operator, for $\alpha$  a complex parameter; $\hat{a}$ and $\hat{a}^{\dagger}$ are the photon annihilation and creation operators respectively and
\begin{equation} 
\hat{S}(\zeta) = \exp\left(-\frac{1}{2}\zeta \hat{a}^{\dagger2} + \frac{1}{2} \zeta^{*}\hat{a}^{2}\right)
\end{equation}
is the squeezing operator with $\zeta = r e^{i\varphi}$; where $\zeta$ is the squeezing parameter;  $r$ and $\varphi$ denote the amplitude and phase of $\zeta$ respectively. The density operator of a thermal radiation field with a heat bath at temperature $T$ can be written as 
\begin{equation}
\hat{\rho}_{\textsubscript{th}} = \frac{1}{1 + \bar{n}_{th}}\sum_{n=0}^{\infty}\left(
\frac{\bar{n}_{th}}{\bar{n}_{th} + 1}\right)^{n}\ket{n}\bra{n},
\end{equation}
where $\bar{n}_{th}$ is the average number of thermal photons and it is given by
\begin{equation}
    \bar{n}_{th} = \frac{1}{\exp\left(\frac{h \nu}{k_B T}\right)-1};
    \label{thermal_avg}
\end{equation}
$k_B$ is Boltzmann constant and $\nu$ is linear frequency of radiation field in Eq. \ref{thermal_avg}. 
The analytic expression for the PCD of SCTS can be written as\cite{PhysRevA.47.4474, PhysRevA.47.4487}

\begin{align}
P(l) =& \bra{l}\hat{\rho}_{\text{SCT}}\ket{l}\\
 =&~ \pi Q(0) \tilde{X}^{l}\sum_{q=0}^{l}\frac{1}{q!}\left(\frac{l}{q}\right)\Big|\frac{|\tilde{Y}|}{2 \tilde{X}}\Big|^{q}\nonumber\\
&\times \big|H_{q}((2Y)^{-1\slash 2} \tilde{Z})\big|^{2},
\end{align}
where $\pi Q(0) = R(0,0)$; $R$ is Glauber's $R$-function\cite{PhysRev.131.2766};

\begin{equation}
R(0,0) = \left[(1 + X)^{2} - |Y|^{2}\right]^{-1\slash 2}\exp\left\{- \frac{(1+X)|Z|^{2} + \frac{1}{2}[Y(Z^{*})^{2} + Y^{*}Z^{2}]}{(1+X)^{2}-|Y|^{2}}\right\},
\end{equation}
where
\begin{align}
X &= \bar{n}_{th} + (2 \bar{n}_{th} + 1) (\sinh r)^{2}, \\
Y &= - (2 \bar{n}_{th} + 1)e^{i\varphi} \sinh r \cosh r,\\
Z &= \alpha  \hspace{0.5cm} \text{(for SCTS)},\\
\end{align}
and
\begin{align}
\tilde{X} &= \frac{X(1 + X) - |Y|^{2}}{(1 + X)^{2} - |Y|^{2}}, \\
\tilde{Y} &= \frac{Y}{(1 + X)^{2} - |Y|^{2}},\\
\tilde{Z} &= \frac{(1 + X)Z + YZ^{*}}{(1 + X)^{2} - |Y|^{2}}.
\end{align}
If we write $\tilde{X}$, $\tilde{Y}$ and $\tilde{Z}$ in terms of $\bar{n}_{th}$ and $r$, we get
\begin{align}
\tilde{X} &= \frac{\bar{n}_{th}(\bar{n}_{th} + 1)}{\bar{n}_{th}^{2} + (\bar{n}_{th} + \frac{1}{2})[1 + \cosh (2r)]}\\
\tilde{Y} &= -\frac{e^{i\varphi}(\bar{n}_{th} + \frac{1}{2})\sinh(2r)}{\bar{n}_{th}^{2} + (\bar{n}_{th} + \frac{1}{2})[1 + \cosh (2r)]}\\
\tilde{Z} &= \frac{Z[\frac{1}{2} + (\bar{n}_{th}+ \frac{1}{2})\cosh r] - Z^{*}e^{i \varphi}(\bar{n}_{th}+ \frac{1}{2}) \sinh 2r}{\bar{n}_{th}^{2} + (\bar{n}_{th} + \frac{1}{2})[1 + \cosh (2r)]} \nonumber\\
\end{align}
and $H_{q}$ is the Hermite polynomial. It is defined as
\begin{equation}
H_{q}(x) = \sum_{j=0}^{\lfloor\frac{ q}{2}\rfloor}\frac{(-1)^{j}q!}{j!(q-2j)!}(2x)^{q-2j}.
\end{equation}
The average number of coherent photons is defined as 
\begin{equation}
    \bar{n}_c = |\alpha|^2,
\end{equation}
and the average number of squeezed photons is defined as
\begin{equation}
    \bar{n}_s = \sinh^2 r.
\end{equation}

\subsection{Atomic state}
The atomic states which are used for our purpose are a Bell state and a Werner state. The Bell state which is used here is
\begin{equation}
\ket{\psi_\textsubscript{AB}} = \frac{1}{\sqrt{2}} (\ket{e_{\text{A}}, g_{\text{B}}} + \ket{g_{\text{A}}, e_{\text{B}}})
\label{bellstate},
\end{equation}
where $|e_{\text{A}} \rangle$ and $|g_{\text{A}} \rangle$ correspond to the excited and ground states of the atom A. Bell states are a set of four maximally entangled two-qubit quantum states that form a fundamental cornerstone in quantum information theory. Their symmetry, purity, and maximal entanglement make them ideal testbeds for understanding decoherence, entanglement dynamics, and the transition between classical and quantum regimes. For the mixed states we consider the Werner-type states which are constructed by mixing the maximally entangled states with the maximally mixed states. Considering the Bell state $|\psi^{-} \rangle = \frac{1}{\sqrt{2}} (|g_{\text{A}},e_{\text{B}}\rangle - |e_{\text{A}},g_{\text{B}}\rangle )$, the Werner state is generated as given below: 
\begin{equation}
    W_\textsubscript{AB} = (1- \eta)\frac{\textit{I}}{4} + \eta\, \ket{\psi^{-}}\bra{\psi^{-}}
    \label{WernerBellstate},
\end{equation}
where $\eta$ is the mixing parameter. When $\eta = 1$ in Eq. (\ref{WernerBellstate}), the state is the maximally entangled Bell state and for $\eta =0$ the state is maximally mixed. In the regime $1/3 \leq \eta \leq 1$, the two qubit Werner state is entangled and for $0 \leq \eta <1/3$ the state is separable. Werner states are physically significant because they represent a bridge between idealized entangled states and noisy, real-world quantum systems. Werner-type mixed states, show non-classical correlations and they can be realized experimentally also by polarization-entangled photon states \cite{PhysRevLett.92.177901}. In Ref. \cite{Czerwinski_2021}, the authors have quantified the entanglement for this two-qubit states. Werner states play a crucial role in quantum information theory, quantum teleportation, etc.\cite{PhysRevLett.84.4236,PhysRevA.66.062312}  Werner states have also been used in noisy quantum channels\cite{PhysRevA.76.052306}. Therefore, it is important to study the entanglement dynamics of the atom-field system for the Werner state.
\subsection{Hamiltonian and time evolution of the system}
The Hamiltonian that describes the interaction between the atoms and the radiation field in this system is given by:
\begin{align}
\hat{H} = \omega\, \hat{\sigma}_{z}^{\text{A}}+\omega\, \hat{\sigma}_{z}^{\text{B}}+ \nu \hat{a}^{\dagger} \hat{a} +\lambda \left(\hat{a}^{\dagger} 
\hat{\sigma}_{-}^{\text{A}}+\hat{a} \hat{\sigma}_{+}^{\text{A}}\right)+ \lambda \left(\hat{a}^{\dagger} \hat{\sigma}_{-}^{\text{B}}+\hat{a} 
\hat{\sigma}_{+}^{\text{B}}\right),
\label{djcmmodel2}
\end{align}
where A and B are the two atoms inside the cavity . The factor $\hat{\sigma}_{z}^{i}$ represents the Pauli matrix in the $z$-basis and $\hat{\sigma}_{+}^{i}$ and $\hat{\sigma}_{-}^{i}$ are the spin raising and lowering operators respectively. The index $i$ represents the atomic label. The photonic operators $\hat{a}$ and $\hat{a}^{\dagger}$ are the photon annihilation and creation operators of the radiation field respectively. The coupling constant is represented by $\lambda$ and it describes the strength of the atom-field interaction with $\omega$ and $\nu$ being the atomic transition frequency and the radiation frequency respectively. Note that there is no interaction between the two atoms, except that they are entangled to start with.

Density operator for the atomic state is 
\begin{equation}
    \hat{\rho}_\textsubscript{AB}(0) = \ket{\psi_\textsubscript{AB}} \bra{\psi_\textsubscript{AB}},
    \label{rho_atom}
\end{equation} 
and initial density operator representing the field states is $\hat{\rho}_{\text{F}}(0)$. So, the density operator for the whole system can be written as 
\begin{align}
    \hat{\rho}_{\text{tot}}(0) =& \hat{\rho}_\textsubscript{AB}(0) \otimes \hat{\rho}_{\text{F}}(0),
    \label{rhotot_ini}
\end{align}
where $\hat{\rho}_{\text{F}}(0)$ are the density operator for the radiation field in the cavity.

The atomic subsystem is restricted to the symmetric (bright) Dicke manifold
\[
\mathcal H_{\mathrm{at}}^{(+)}=\mathrm{span}\left\{
|gg\rangle,\;|\psi_\textsubscript{AB}\rangle,\;|ee\rangle
\right\},
\qquad
|\psi_\textsubscript{AB}\rangle=\frac{|eg\rangle+|ge\rangle}{\sqrt{2}} .
\]
In the interaction picture and under the resonance condition
\(\omega_c=\omega_0\),
the Hamiltonian reads
\[
\hat{H}_I=\hbar g\left(\hat{a}\,\hat{J}^+ + \hat{a}^\dagger \hat{J}^-\right),
\qquad
\hat{J}^\pm=\hat{\sigma}_A^\pm+\hat{\sigma}_B^\pm .
\]
In the ordered basis
\(\{|gg\rangle,|\psi_\textsubscript{AB}\rangle, |ee\rangle\}\),
the Hamiltonian takes the matrix form
\begin{equation}
\hat{H}_I
=
\hbar
\begin{pmatrix}
0 & \sqrt{2}g\,\hat{a}^\dagger & 0 \\
\sqrt{2}g\,\hat{a} & 0 & \sqrt{2}g\,\hat{a}^\dagger \\
0 & \sqrt{2}g\,\hat{a} & 0
\end{pmatrix}.
\end{equation}
The unitary time-evolution operator is
\[
\hat{U}(t)=\exp\!\left(-\frac{i}{\hbar}\hat{H}_I t\right).
\]
Define the field number operator
\[
\hat{N}=\hat{a}^\dagger \hat{a}
\]
and the single nonzero operator-valued collective Rabi frequency
\[
\hat{\Omega} \equiv \hat{\Omega}_2 = \sqrt{2g^2 (\hat{N}+1)} .
\]
The exact evolution operator can then be written as
\begin{equation}
\hat{U}(t)=
\begin{pmatrix}
\dfrac{\hat{N} + (\hat{N}+1)\cos(\hat{\Omega} t)}{2\hat{N}+1} &
-i\,\dfrac{\sqrt{2(\hat{N}+1)}}{\hat{\Omega}}\,
\sin(\hat{\Omega} t)\,\hat{a}^\dagger &
\dfrac{\sqrt{2\hat{N}(\hat{N}+1)}}{2\hat{N}+1}
\bigl[\cos(\hat{\Omega} t)-1\bigr]
\\[2mm]
-i\,\dfrac{\sqrt{2(\hat{N}+1)}}{\hat{\Omega}}\,
\sin(\hat{\Omega} t)\,\hat{a} &
\cos(\hat{\Omega} t) &
-i\,\dfrac{\sqrt{2\hat{N}}}{\hat{\Omega}}\,
\sin(\hat{\Omega} t)\,\hat{a}^\dagger
\\[2mm]
\dfrac{\sqrt{2\hat{N}(\hat{N}+1)}}{2\hat{N}+1}
\bigl[\cos(\hat{\Omega} t)-1\bigr] &
-i\,\dfrac{\sqrt{2\hat{N}}}{\hat{\Omega}}\,
\sin(\hat{\Omega} t)\,\hat{a} &
\dfrac{\hat{N}+1 + \hat{N}\cos(\hat{\Omega} t)}{2\hat{N}+1}
\end{pmatrix}.
\end{equation}
All trigonometric functions are defined as operator functions of \(\hat{N}\)
via the spectral theorem. The exact time-evolved density operator is
\[
\hat{\rho}_{\text{tot}}(t)=\hat{U}(t)\,\hat{\rho}(0)\,\hat{U}^\dagger(t).
\]
It can be expressed as
\begin{equation}
\hat{\rho}_{\text{tot}}(t)
=
\sum_{i,j\in\{gg,\psi_\textsubscript{AB},ee\}}
|i\rangle\langle j|\otimes\hat{\rho}^{(f)}_{ij}(t),
\label{rho_t}
\end{equation}
where the nonvanishing field operators are
\[
\hat{\rho}^{(f)}_{\psi_\textsubscript{AB}\psi_\textsubscript{AB}}(t)
=
\cos(\hat{\Omega} t)\,\hat{\rho}_f\,\cos(\hat{\Omega} t),
\]

\[
\hat{\rho}^{(f)}_{gg\,gg}(t)
=
\frac{2(\hat{N}+1)}{\hat{\Omega}^2}\sin^2(\hat{\Omega} t)\;
\hat{a}^\dagger\hat{\rho}_f \hat{a},
\]

\[
\hat{\rho}^{(f)}_{ee\,ee}(t)
=
\frac{2 \hat{N}}{\hat{\Omega}^2}\sin^2(\hat{\Omega} t)\;
\hat{a}\,\hat{\rho}_f \hat{a}^\dagger,
\]

\[
\hat{\rho}^{(f)}_{gg\,\psi_\textsubscript{AB}}(t)
=
-i\,\frac{\sqrt{2(\hat{N}+1)}}{\hat{\Omega}}\sin(\hat{\Omega} t)\, \hat{a}^\dagger \hat{\rho}_f,
\]

\[
\hat{\rho}^{(f)}_{\psi_\textsubscript{AB}\,ee}(t)
=
-i\,\frac{\sqrt{2 \hat{N}}}{\hat{\Omega}}\sin(\hat{\Omega} t)\, \hat{\rho}_f \hat{a}^\dagger,
\]
with the remaining coherences obtained by Hermitian conjugation.

\subsection{Computing entanglement for different subsystems}

Although, an analytical expression for the density operator corresponding to a squeezed coherent thermal state (SCTS) is highly involved and no closed-form expansion in the Fock basis is known, the entanglement dynamics of the system can still be reliably and robustly studied through numerical methods. In particular, by evolving the total density matrix and performing partial traces over the appropriate subsystems, one can investigate the entanglement between atoms, fields, and hybrid subsystems. 

The expression in \ref{rho_t} is the most general form of the time-evolved density matrix. It is written in terms of the initial field density operator and operator functions of the number operator. For example, to find the reduced density matrix of the atoms $\hat{\rho}_{\text{atoms}}(t)$, we must trace over the field degrees of freedom:
\begin{equation}
\hat{\rho}_{\text{atoms}}(t)=\text{Tr}_{\text{field}}[\hat{\rho}_{\text{tot}}(t)].
\end{equation}
We can do similar operations for other subsystems also.

\subsection{Entanglement measures}
To characterize the dynamics of entanglement, we need to measure the entanglement in the system. In this work, we investigate the dynamics of the bipartite entanglements like the atom-atom entanglement, and atom-field entanglement. The atom-atom entanglement can be conclusively measured using concurrence defined in \cite{wootters2001entanglement}
\begin{equation}
C_{\text{AB}} = \text{max}\{0, \Lambda_{1} - \Lambda_{2}-\Lambda_{3}-\Lambda_{4}\},
\end{equation}
where $\Lambda_{i} (i = 1, 2, 3, 4)$ are the decreasingly ordered square roots of the eigenvalues of the matrix 
$\hat{\rho} \left(\hat{\sigma}_{y}^{\text{A}}\otimes \hat{\sigma}_{y}^{\text{B}}\right) \hat{\rho}^{*}\\
\left(\hat{\sigma}_{y}^{\text{A}} \otimes \hat{\sigma}_{y}^{\text{B}}\right)$ and $\hat{\rho}$ is the two qubit atom-atom reduced density matrix. The value of concurrence lies in the range $0 \leq C \leq 1$, where $C=0$ implies a separable state and $C=1$ denotes a maximally entangled state. Though concurrence can be used to compute entanglement in both pure and mixed states, if the system is $2 \otimes 2$ dimensions. So, for higher dimensional systems we need to use other measures. In particular, when we consider the atom-field subsystem we are looking at $2 \otimes \infty$ and bipartite continuous variable systems. For these systems it is convenient to use the negativity \cite{wei2003maximal} which is defined as 
\begin{equation}
N(\rho)=\sum_{k}\Big( |\xi_{k}|-\xi_{k} \Big)/2, 
\end{equation}
where $\xi_{k}$ are the eigenvalues of $\hat{\rho}^{\text{PT}}$, the partial transpose of the density matrix, i.e., the matrix which is transposed with respect to any one of the subsystems.

\section{Effects of thermal and squeezed photons on entanglement}

\begin{figure}
    \centering
    \includegraphics[scale = 0.45]{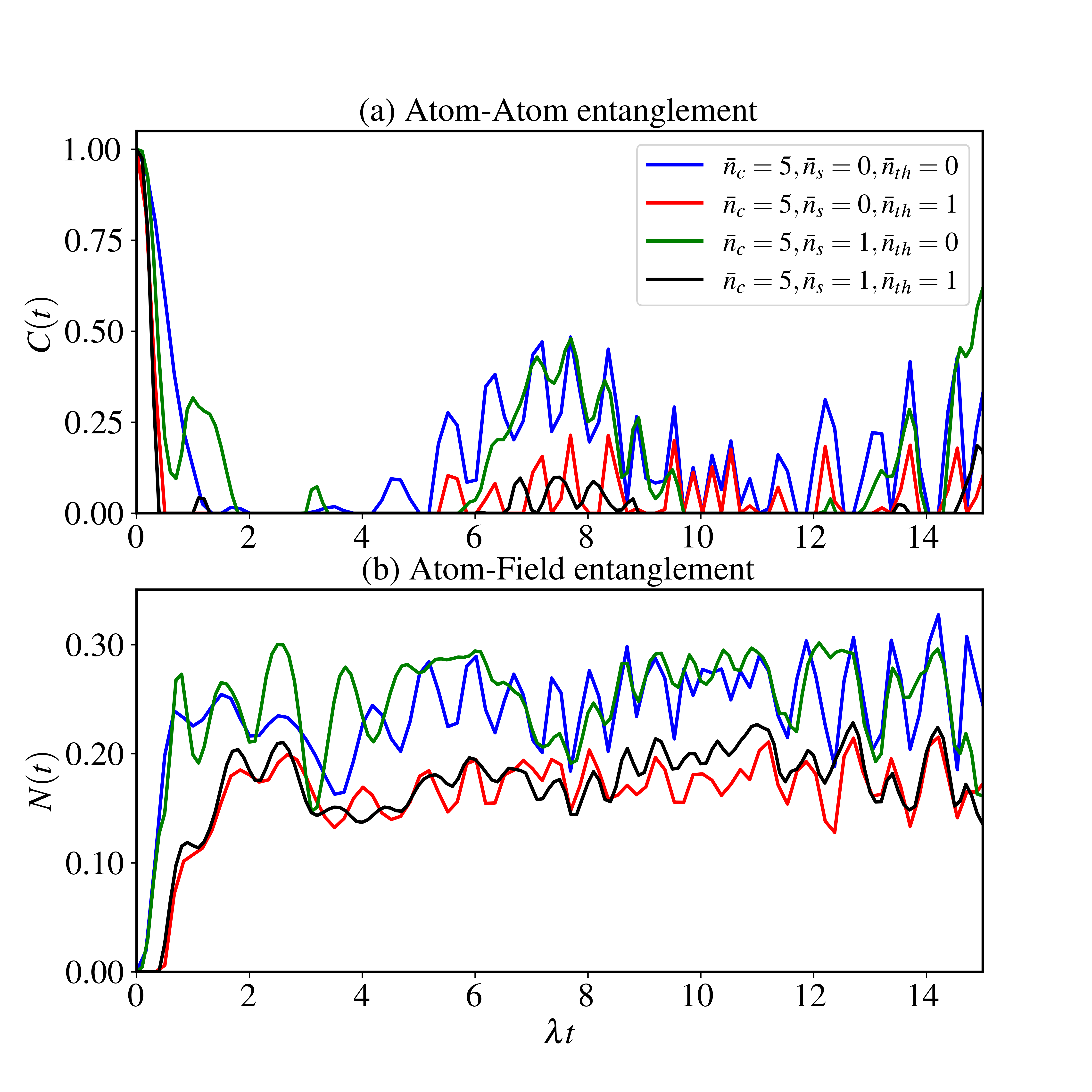}
    \caption{Effects of squeezed photons and thermal photons on entanglement dynamics with atoms in Bell state. Here, (a) blue curve $\Rightarrow  \bar{n}_c = 5, \bar{n}_s = 0, \bar{n}_{th} = 0$, (b) green curve $\Rightarrow  \bar{n}_c = 5, \bar{n}_s = 0, \bar{n}_{th} = 1$, (c) red curve $\Rightarrow  \bar{n}_c = 5, \bar{n}_s = 1, \bar{n}_{th} = 0$ and  (d) black curve $\Rightarrow  \bar{n}_c = 5, \bar{n}_s = 1, \bar{n}_{th} = 1$.}
    \label{fig_1}
\end{figure}

\begin{figure}
    \centering
    \includegraphics[scale = 0.3]{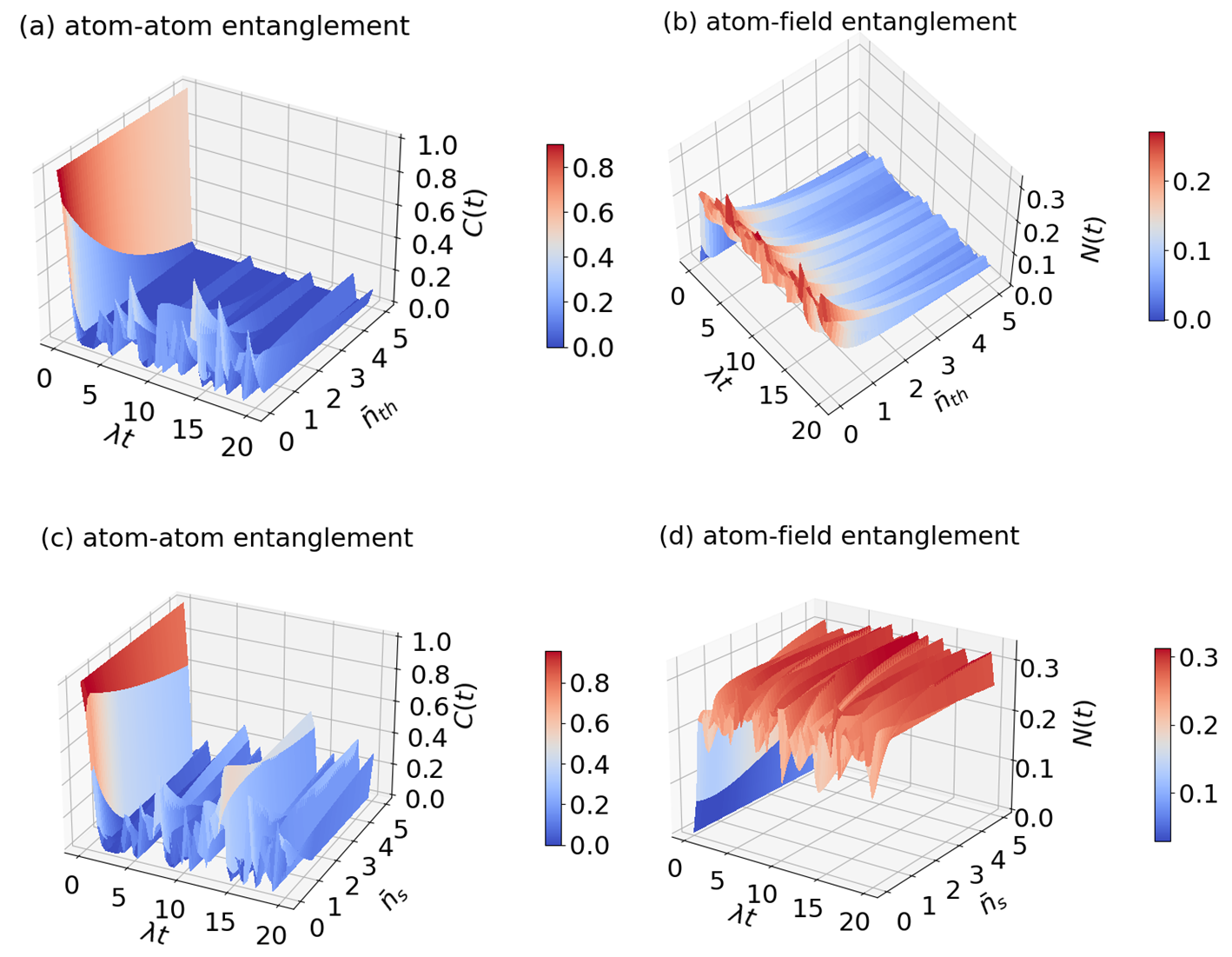}
    \caption{(a), (b) represent 3D plots for $C(t), N(t)$ vs $\bar{n}_{th}$ and $\lambda t$; (c), (d) represent 3D plots for $C(t), N(t)$  vs $\bar{n}_{s}$ and $\lambda t$ for the atoms in Bell state. Here, in Fig. 2(a), 2(b) $\bar{n}_c = 2, \bar{n}_s = 0$ and $\bar{n}_{th}$ is varied; in Figs. 3(c), 3(d) $\bar{n}_c = 2, \bar{n}_{th} = 0$ and $\bar{n}_{s}$ is varied.}
    \label{fig_2}
\end{figure}

\noindent{\textit{\textbf{For atoms in Bell state:}}}\\
The effects of the thermal photon number $\bar{n}_{th}$ and the squeezed photon number $\bar{n}_s$ on the entanglement dynamics and atomic inversion for a squeezed coherent thermal state (SCTS), with the atoms initially prepared in a Bell state, is illustrated in Figs.~\ref{fig_1} and \ref{fig_2}.  

For a purely coherent field ($\bar{n}_{th}=\bar{n}_s=0$, blue curve), the atomic concurrence $C(t)$ initially attains its maximum value and then rapidly decreases due to atom--field interaction. As time evolves, entanglement sudden death (ESD) appears in the atom--atom entanglement, followed by partial revivals at later times. This behaviour reflects the well-known collapse--revival structure of the Jaynes--Cummings-type dynamics. In contrast, the atom--field entanglement, quantified by the negativity $N(t)$, starts from zero since the atoms and field are initially separable. Subsequently, $N(t)$ grows rapidly and exhibits oscillatory behaviour without any ESD, highlighting a qualitative difference between atom--atom and atom--field entanglement dynamics.

When a single thermal photon is added to the coherent field ($\bar{n}_{th}=1$, red curve), the concurrence $C(t)$ is strongly affected. In particular, several smaller revival peaks disappear, leading to a significant increase in the duration of ESD intervals. A small ESD also emerges in the atom--field entanglement $N(t)$, accompanied by a substantial reduction in its oscillation amplitude. This degradation of both atom--atom and atom--field entanglement can be attributed to thermal photons acting as ``classical noise'', which introduces incoherent fluctuations and washes out quantum correlations in the system.

In contrast, when a squeezed photon is added to the coherent field ($\bar{n}_s=1$, green curve in Fig.~\ref{fig_1}), the dynamics exhibit a markedly different behaviour. Both $C(t)$ and $N(t)$ increase appreciably, and the number as well as the duration of ESDs are reduced. This enhancement arises from the fact that squeezed photons represent ``quantum noise'' with reduced fluctuations in one quadrature, which can induce and amplify quantum correlations. Nevertheless, the entanglement enhancement due to squeezing is weaker than the entanglement suppression caused by thermal photons, indicating that classical noise is more detrimental than the beneficial effects of moderate squeezing.

When both thermal and squeezed photons are simultaneously present ($\bar{n}_{th}=\bar{n}_s=1$, black curve), the resulting dynamics reflect a competition between these two opposing effects. As previously discussed in Ref.~\cite{mandal2023atomic} for a single-atom system, thermal photons tend to destroy entanglement, whereas squeezed photons attempt to restore it. In the present two-atom scenario, squeezing partially recovers the lost entanglement and enhances the amplitudes of both $C(t)$ and $N(t)$. However, the thermal contribution dominates, and the overall dynamics remain closer to those of a thermal coherent state.

Figure~\ref{fig_2} presents three-dimensional plots of $C(t)$ and $N(t)$ as functions of $\bar{n}_{th}$ and $\bar{n}_s$, with $\bar{n}_c=2$ and $\bar{n}_{th}, \bar{n}_s$ varied from $0$ to $5$. This allows a comparison between two regimes: (i) $\bar{n}_c > \bar{n}_{th}, \bar{n}_s$ and (ii) $\bar{n}_c < \bar{n}_{th}, \bar{n}_s$. From Figs.~\ref{fig_2}(a) and (b), it is evident that for each value of $\bar{n}_{th}$, $C(t)$ initially starts from a maximum. In the regime $\bar{n}_c > \bar{n}_{th}$, the concurrence exhibits multiple revival peaks with relatively short ESD intervals. Conversely, when $\bar{n}_c < \bar{n}_{th}$, many revival peaks vanish, the overall amplitude decreases, and the ESD durations become significantly longer. The atom--field entanglement $N(t)$ is also strongly suppressed with increasing thermal photon number, underscoring the destructive role of thermal noise.\\

Figures~\ref{fig_2}(c) and (d) demonstrate that increasing the squeezed photon number $\bar{n}_s$ enhances the amplitudes of both $C(t)$ and $N(t)$ in both regimes, i.e., $\bar{n}_c < \bar{n}_s$ and $\bar{n}_c > \bar{n}_s$. This confirms that squeezing acts as a robust resource for inducing and strengthening entanglement in the system, partially counteracting the deleterious effects of thermal fluctuations.\\

\begin{figure}
    \centering
    \includegraphics[scale = 0.45]{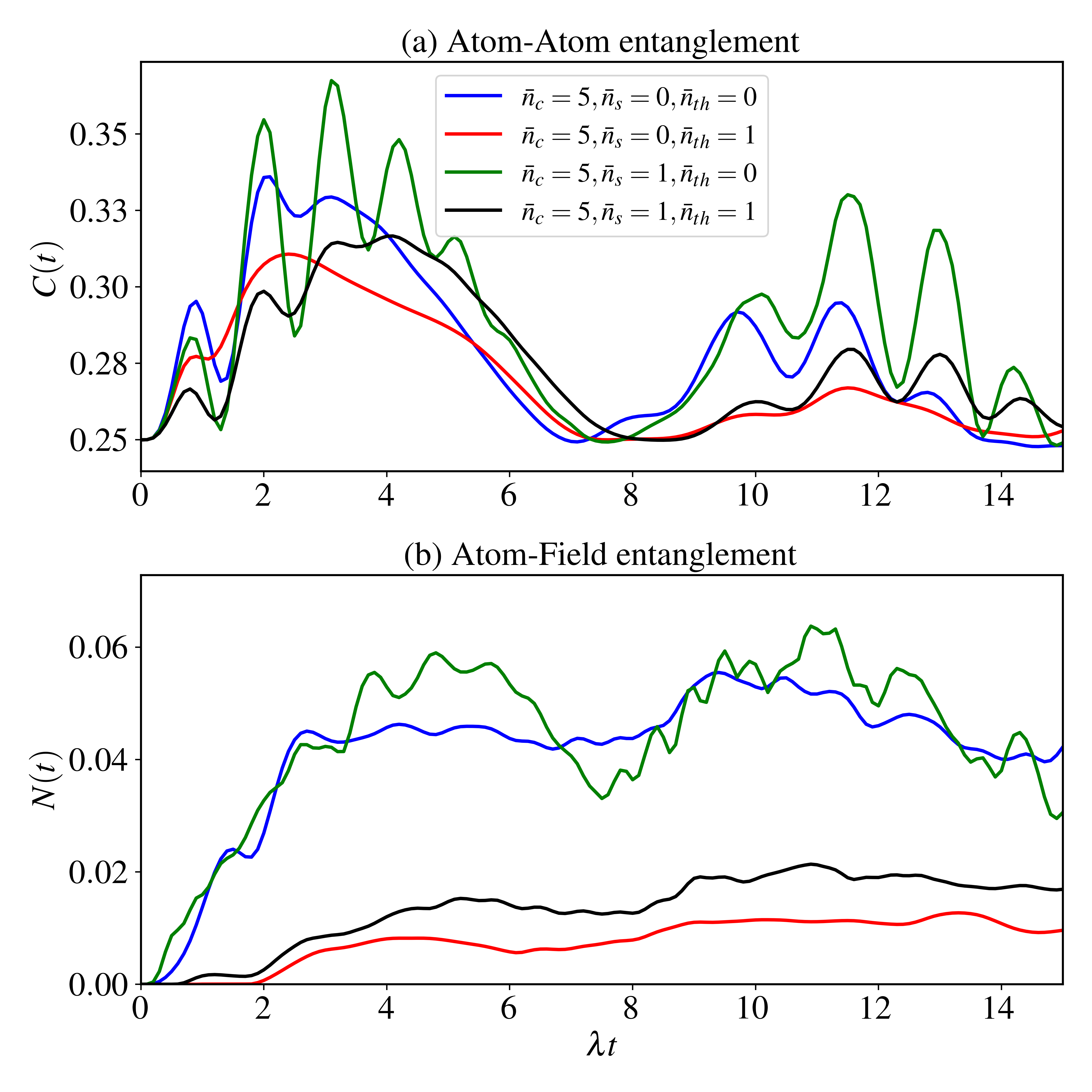}
    \caption{Effects of squeezed photons and thermal photons on entanglement dynamics with atoms in Werner state. Here, (a) blue curve $\Rightarrow  \bar{n}_c = 5, \bar{n}_s = 0, \bar{n}_{th} = 0$, (b) green curve $\Rightarrow  \bar{n}_c = 5, \bar{n}_s = 0, \bar{n}_{th} = 1$, (c) red curve $\Rightarrow  \bar{n}_c = 5, \bar{n}_s = 1, \bar{n}_{th} = 0$ and  (d) black curve $\Rightarrow  \bar{n}_c = 5, \bar{n}_s = 1, \bar{n}_{th} = 1$.}
    \label{fig_3}
\end{figure}

\noindent{\textit{\textbf{For atoms in Werner state:}}}\\
The effects of the thermal photon number $\bar{n}_{th}$ and the squeezed photon number $\bar{n}_{s}$ on the dynamics of the atomic concurrence $C(t)$ and the atom--field negativity $N(t)$, with the atoms initially prepared in a Werner state, are shown in Fig.~\ref{fig_3}. In contrast to the Bell-state case, it is observed that $C(t)$ does not exhibit entanglement sudden death (ESD) for any of the considered combinations of $\bar{n}_c$, $\bar{n}_{th}$, and $\bar{n}_s$. Although the addition of thermal photons significantly suppresses the amplitude of $C(t)$, the atom--atom entanglement never vanishes completely. This behaviour highlights an important qualitative difference between the two initial states: while Bell-state entanglement is highly fragile and prone to ESD, the mixed nature of the Werner state renders the atomic entanglement more robust against decoherence induced by thermal noise.

Despite this robustness in $C(t)$, the atom--field entanglement $N(t)$ is considerably weaker for atoms in a Werner state compared to the Bell-state case. As seen in Fig.~\ref{fig_3}, $N(t)$ displays small ESD intervals at the early stages of the dynamics, which become more pronounced with the inclusion of thermal photons. This indicates that, although atomic entanglement is preserved, the ability of the atoms to establish strong correlations with the field is reduced due to the initial mixedness of the atomic state. Consequently, $C(t)$ and $N(t)$ exhibit qualitatively opposite trends when comparing Bell and Werner initial states.

\begin{figure}
    \centering
    \includegraphics[scale = 0.45]{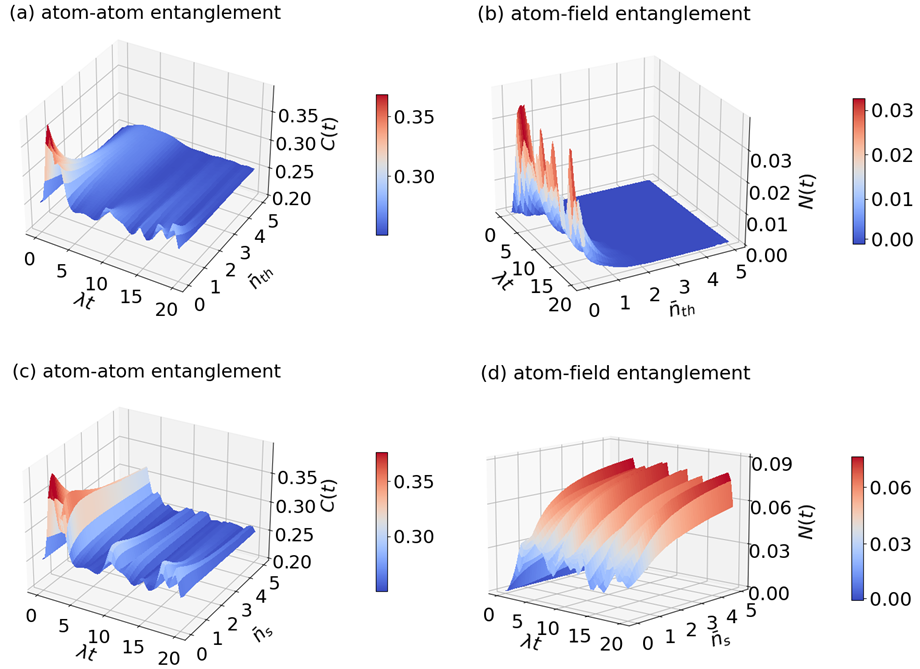}
    \caption{(a), (b) represent 3D plots for $C(t), N(t)$ vs $\bar{n}_{th}$ and $\lambda t$; (c), (d) represent 3D plots for $C(t), N(t)$  vs $\bar{n}_{s}$ and $\lambda t$ for the atoms in Werner state ($\eta = 0.5$).  Here, in Fig. 4(a), 4(b) $\bar{n}_c = 2, \bar{n}_s = 0$ and $\bar{n}_{th}$ is varied; in Fig. 4(c), 4(d) $\bar{n}_c = 2, \bar{n}_{th} = 1$ and $\bar{n}_{s}$ is varied.}
    \label{fig_4}
\end{figure}
Figures~\ref{fig_4}(a) and (b) present three-dimensional plots illustrating the dependence of $C(t)$ and $N(t)$ on the thermal photon number $\bar{n}_{th}$ for atoms in a Werner state. As shown in Fig.~\ref{fig_4}(a), increasing $\bar{n}_{th}$ leads to a systematic reduction in the amplitude of $C(t)$, confirming the detrimental role of thermal photons on atomic entanglement. However, even for large thermal photon numbers, $C(t)$ remains nonzero, further emphasizing the absence of ESD in this case. In contrast, the atom--field entanglement $N(t)$ starts with a relatively small amplitude and decreases sharply with increasing $\bar{n}_{th}$, indicating that thermal noise strongly suppresses atom--field correlations.

Figure~\ref{fig_4}(c) demonstrates the effect of squeezed photons on the entanglement dynamics. For the regime $\bar{n}_c > \bar{n}_{s}$, the peak values of $C(t)$ decrease with increasing $\bar{n}_{s}$, although atomic entanglement is never completely lost. Meanwhile, the atom--field entanglement $N(t)$ increases monotonically with $\bar{n}_{s}$, illustrating that squeezing enhances atom--field correlations even when the atoms are initially in a mixed state. Nevertheless, it is important to note that the overall magnitude of $N(t)$ remains significantly smaller than that obtained for atoms initially prepared in a Bell state, underscoring the limiting role of initial mixedness in the generation of atom--field entanglement.

\subsection{Wigner distribution functions }

In this subsection, we analyze the Wigner distribution function $W(\alpha)$ of the squeezed coherent thermal state (SCTS) for atoms initially prepared in Bell and Werner states. The Wigner function provides a quasiprobability representation of the radiation field in phase space and serves as a powerful indicator of nonclassicality, with the appearance of negative regions signaling intrinsically quantum features. $W(\alpha)$ is defined as\cite{HILLERY1984121, schleich, agarwal2013}
\begin{equation}
W(\alpha)=\frac{1}{\pi^{2}}\int d^{2}\beta\enspace \text{Tr}[\hat{\rho}\hat{D}(\beta)]\exp(\beta^{*}\alpha-\beta\alpha^{*}).
\end{equation}
The density operator $\hat{\rho}$ for SCTS is given in Eq. (\ref{rho_sct}).\\

\begin{figure}
    \centering
    \includegraphics[scale = 0.45]{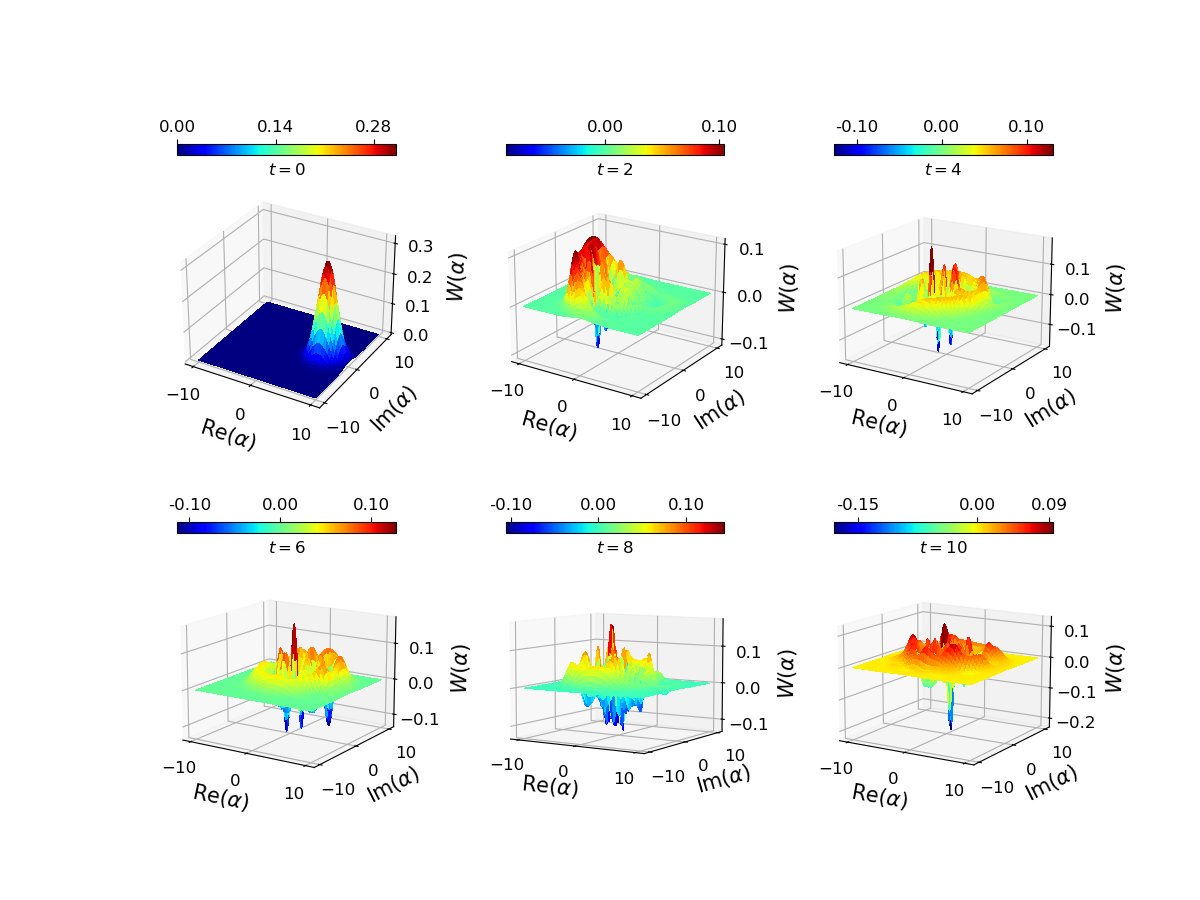}
    \caption{Wigner functions for SCTS in JCM with atoms in a Bell state. Here the other parameters are $\bar{n}_c = 5.0, \bar{n}_s = 0.0, \bar{n}_{th}=0.0$ and $\eta = \frac{\pi}{4}$.}
    \label{wig_jcm_scts_bell1}
\end{figure}

\begin{figure}
    \centering
    \includegraphics[scale = 0.45]{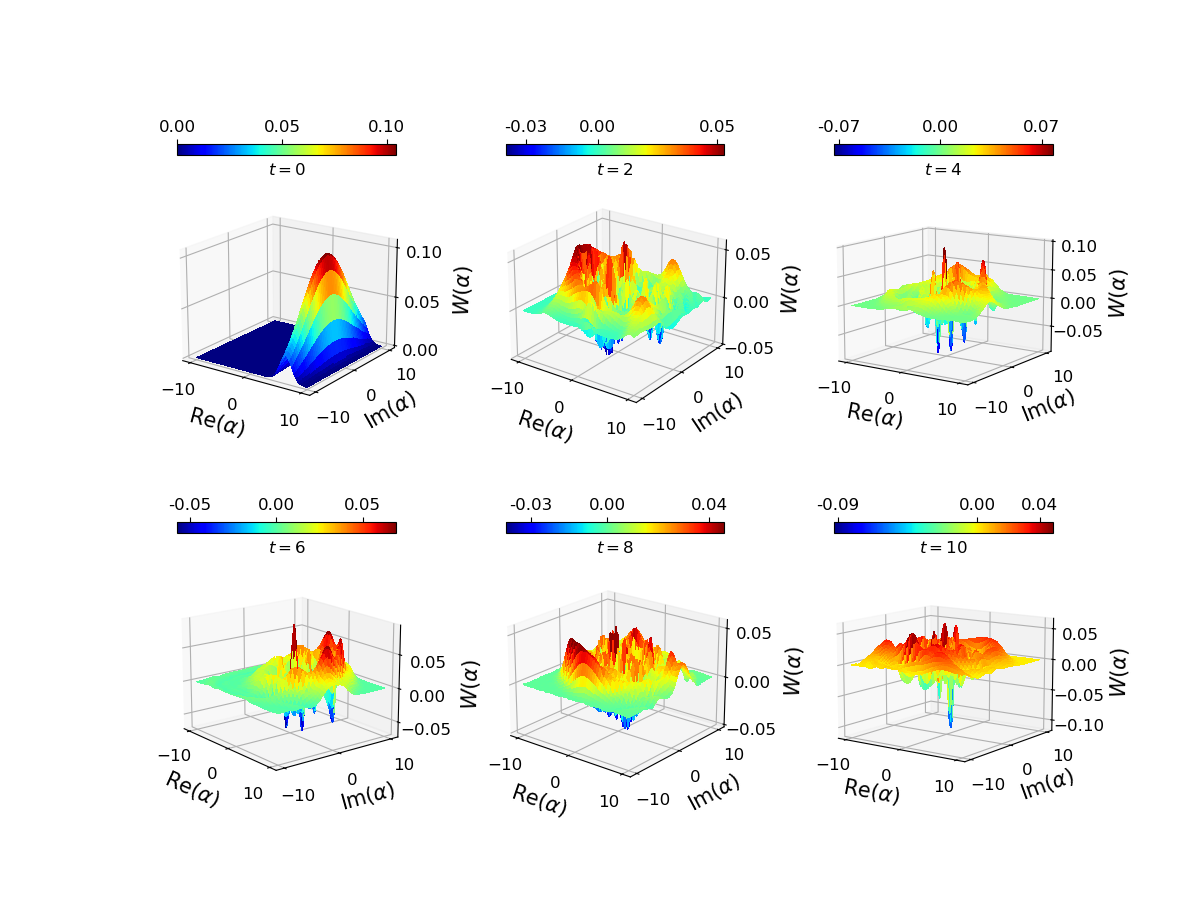}
    \caption{Wigner functions for SCTS in JCM with atoms in a Bell state. Here the other parameters are $\bar{n}_c = 5.0, \bar{n}_s = 1.0, \bar{n}_{th} = 1.0$ and $\theta = \frac{\pi}{4}$.}
    \label{wig_jcm_scts_bell2}
\end{figure}

\noindent \textbf{\textit{For atoms in Bell state:}}\\
The Wigner distribution functions for SCTS interacting with atoms initially in a Bell state are shown in Figs.~\ref{wig_jcm_scts_bell1} and \ref{wig_jcm_scts_bell2}. From Fig.~\ref{wig_jcm_scts_bell1}, it is evident that at $t=0$, the Wigner function exhibits a Gaussian profile centered in phase space, as expected for a coherent radiation field. However, due to the atom--field interaction, $W(\alpha)$ develops negative regions at later times (see Fig.~\ref{wig_jcm_scts_bell1} at $t=2$), indicating the onset of nonclassicality in the radiation field.

As time progresses, the number and depth of negative peaks increase (see Fig.~\ref{wig_jcm_scts_bell1} at $t=4,6,8$), reflecting the continuous transfer of quantum correlations from the atomic subsystem to the field. At $t=10$, the amplitude of the negative peaks is nearly doubled compared to earlier times, demonstrating that the radiation field becomes increasingly nonclassical as a result of sustained interaction with atoms prepared in a maximally entangled pure state.

Figure~\ref{wig_jcm_scts_bell2} shows the Wigner functions for SCTS with $\bar{n}_c=5.0$, $\bar{n}_s=1.0$, and $\bar{n}_{th}=1.0$. The initial Wigner function is positive over the entire phase space and displays a stretched Gaussian profile due to the combined effects of squeezing and thermal photons. At $t=2$, pronounced negative peaks emerge across phase space, indicating that the inclusion of both squeezed and thermal photons enhances the generation of nonclassical features through atom--field interaction. Although the overall amplitude of $W(\alpha)$ is reduced to nearly half of its initial value, the persistence of negative regions confirms the robustness of nonclassicality. As time increases ($t=4,6,8$), the shape of $W(\alpha)$ evolves while the magnitude of the negative peaks remains nearly unchanged. At $t=10$, the depth of the negative regions increases significantly, signifying further enhancement of nonclassical behavior.\\

\begin{figure}
    \centering
    \includegraphics[scale = 0.45]{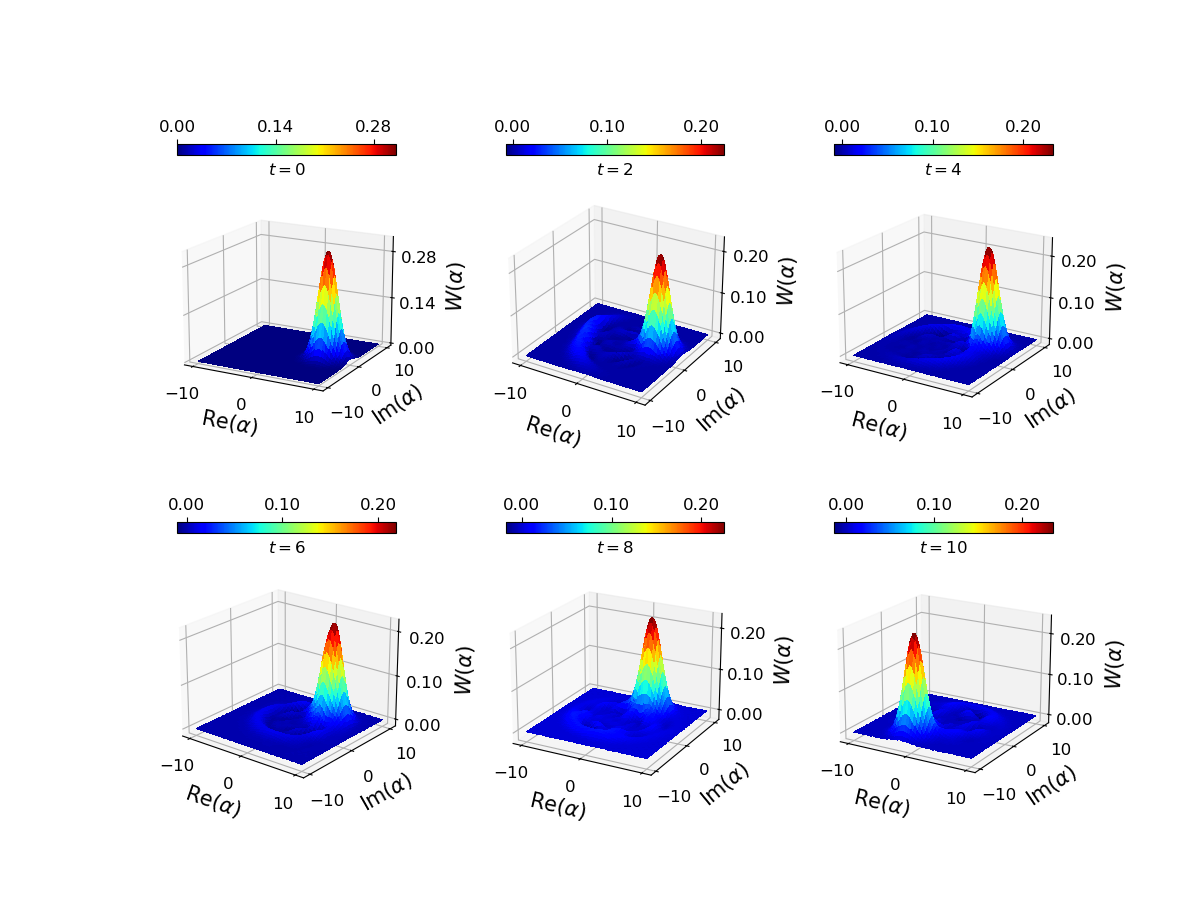}
    \caption{Wigner functions for SCTS in JCM with atoms in a Werner state. Here, the other parameters are $\bar{n}_c = 5.0, \bar{n}_s = 0.0, \bar{n}_{th}=0.0$ and $\eta = 0.5$.}
    \label{wig_jcm_scts_werner1}
\end{figure}

\begin{figure}
    \centering
    \includegraphics[scale = 0.45]{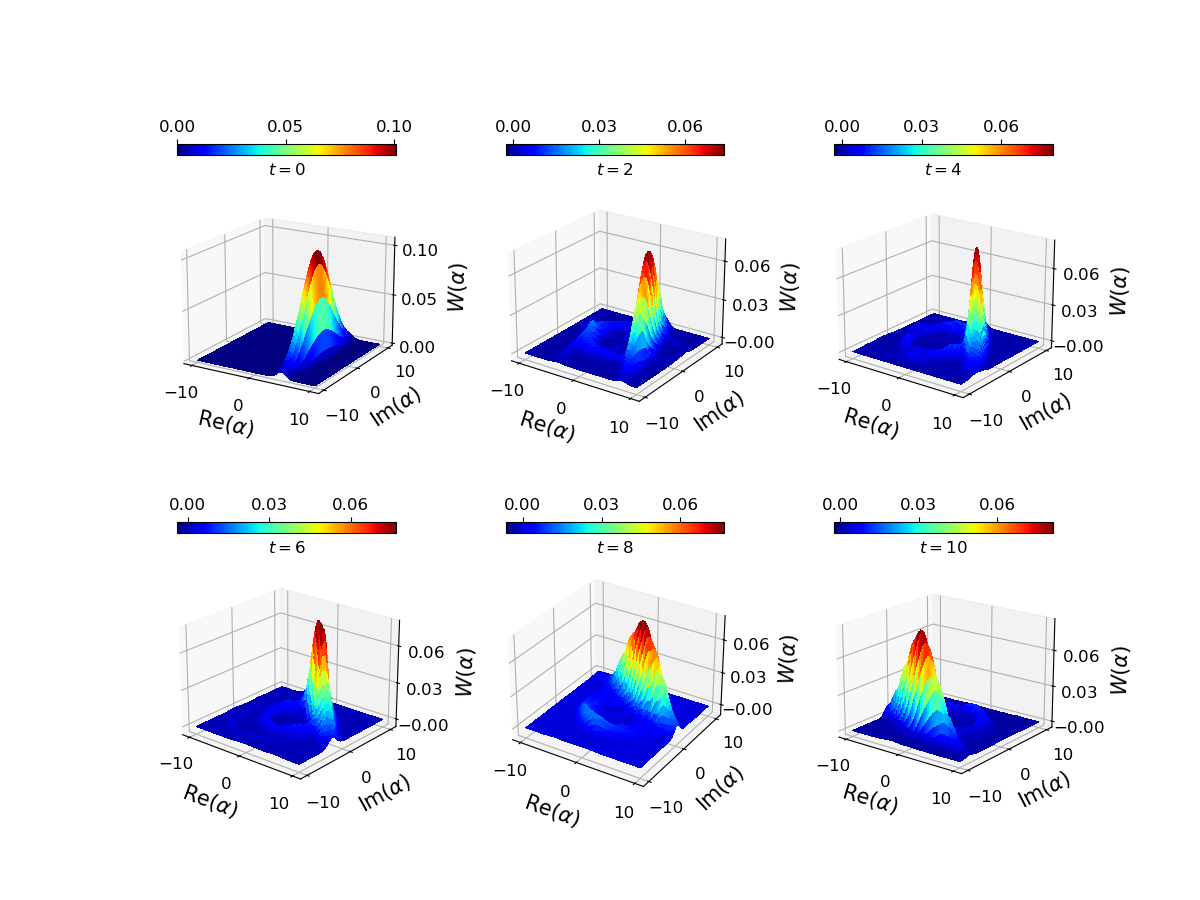}
    \caption{Wigner functions for SCTS in JCM with atoms in a Werner state. Here, the other parameters are $\bar{n}_c = 5.0, \bar{n}_s = 1.0, \bar{n}_{th}=1.0$ and $\eta = 0.5$.}
    \label{wig_jcm_scts_werner2}
\end{figure}

\noindent \textbf{\textit{For atoms in Werner state:}}\\
The Wigner distribution functions for SCTS with atoms initially prepared in a Werner state are presented in Figs.~\ref{wig_jcm_scts_werner1} and \ref{wig_jcm_scts_werner2}. From Fig.~\ref{wig_jcm_scts_werner1}, it is observed that for $\bar{n}_s=0$ and $\bar{n}_{th}=0$, the Wigner function remains positive throughout the entire phase space for all considered times. Any negligible negativity appearing in the color scale can be attributed to numerical artifacts rather than genuine nonclassicality. This behavior indicates that the mixedness of the Werner state suppresses the generation of nonclassical features in the radiation field, effectively preserving the classical nature of the initial coherent state.

When both thermal and squeezed photons are included ($\bar{n}_s=1$, $\bar{n}_{th}=1$), the Wigner function remains positive at all times, as shown in Fig.~\ref{wig_jcm_scts_werner2}. Although the distribution undergoes displacement and deformation in phase space during the evolution, its overall shape remains largely intact and free from negative regions. This suggests that, despite the presence of squeezing, the initial mixedness of the atomic Werner state significantly limits the transfer of quantum correlations to the field.

Overall, the Wigner distribution analysis reveals that the initial atomic state plays a crucial role in determining the quantum character of the radiation field. Atoms prepared in a pure entangled Bell state strongly induce nonclassicality in the field, whereas atoms in a mixed Werner state largely inhibit the emergence of negative regions in the Wigner function. This highlights a direct correspondence between atomic entanglement purity and the generation of nonclassical radiation.

\section{Effects of Ising-type interaction}

In this section, we investigate the influence of an Ising-type interaction between the two atoms on the entanglement dynamics of the atom--atom and atom--field subsystems. Ising-type couplings have attracted considerable attention in the context of double Jaynes--Cummings models (DJCMs), as they introduce an additional interaction channel that can modify the generation and robustness of quantum correlations. For instance, Ghoshal \textit{et al.}~\cite{PhysRevA.101.053805} studied the entanglement dynamics of a quenched disordered DJCM in the presence of Ising-type interaction, while Pandit \textit{et al.}~\cite{Pandit_2018} analyzed the role of such interactions in ordered DJCMs. More recently, Laha~\cite{laha2023dynamics} investigated the effects of Ising-type coupling on entanglement dynamics for vacuum, coherent, and thermal radiation fields, and Sadiek \textit{et al.}~\cite{e23050629} examined the time evolution and asymptotic behavior of entanglement in systems incorporating both dipole--dipole and Ising-type interactions.

Including the Ising-type interaction, the total Hamiltonian of the system becomes
\begin{equation}
\hat{H}_{\text{IS}} = \hat{H} + J_{z}\, \hat{\sigma}^{\text{A}}_{z} \otimes \hat{\sigma}^{\text{B}}_{z},
\label{ising_int}
\end{equation}
where the second term represents the Ising-type interaction between the two atoms, and $\hat{H}$ is the Hamiltonian defined in Eq.~\ref{djcmmodel2}. The Ising interaction introduces a state-dependent energy shift without directly exchanging excitations between the atoms.\\

\begin{figure}
    \centering
    \includegraphics[scale = 0.42]{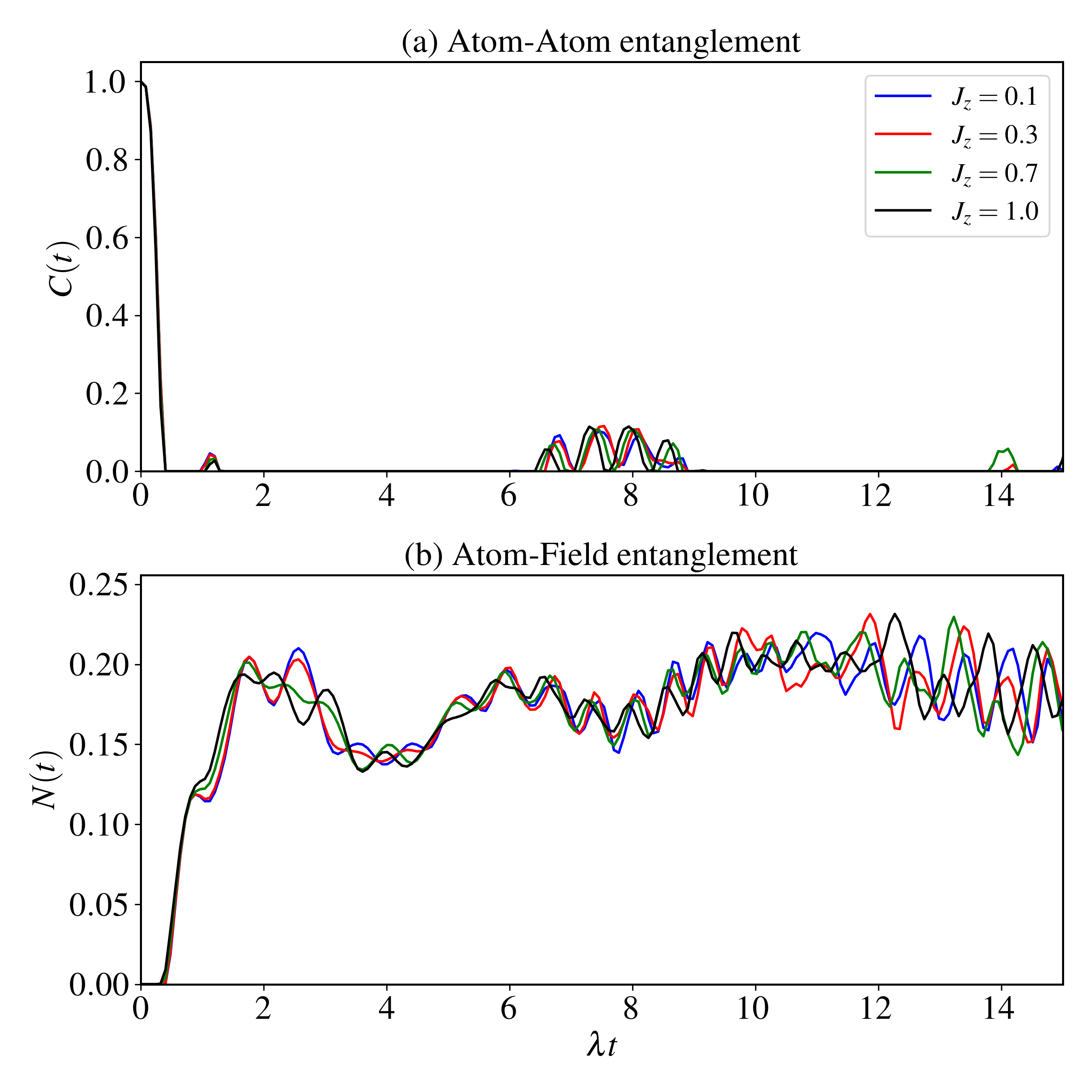}
    \caption{Effects of Ising-type interaction on entanglement dynamics with atoms in Bell state and field is in SCTS. The parameters used in this plot are $\bar{n}_c = 5, \bar{n}_{th} = 1, \bar{n}_s = 1$ and  $J_{z} = 0.1, 0.3, 0.7, 1.0$. Blue curve $\Rightarrow J = 0.1$, red curve $\Rightarrow J = 0.3$, green curve $ \Rightarrow J = 0.7$ and black curve $\Rightarrow J = 1.0$. }
    \label{fig_9}
\end{figure}

\noindent{\textit{\textbf{Atoms in Bell state:}}}\\
The effects of the Ising-type interaction on the entanglement dynamics for atoms initially prepared in a Bell state are shown in Fig.~\ref{fig_9}. The system parameters are chosen as $\bar{n}_c = 5$, $\bar{n}_{th} = 1$, $\bar{n}_s = 1$, with the Ising coupling strength taking the values $J_{z} = 0.1, 0.3, 0.7,$ and $1.0$. As seen in Fig.~\ref{fig_9}(a), the atomic concurrence $C(t)$ exhibits only marginal changes upon increasing $J_z$ when compared to the case without Ising interaction (see the black curve in Fig.~\ref{fig_1}(a)). 

In contrast to the double Jaynes--Cummings model, where the Ising interaction was shown to suppress entanglement sudden death (ESD) and enhance the concurrence amplitude~\cite{Mandal_2024}, such pronounced effects are absent in the present single Jaynes--Cummings configuration. This can be attributed to the fact that the Ising term induces only a relative phase between the atomic states and does not facilitate excitation exchange, thereby having a limited impact on entanglement generation in the single-cavity setup. A similar weak dependence on $J_z$ is observed for the atom--field entanglement $N(t)$, which shows only slight quantitative variations without any qualitative modification of the dynamics.\\

\begin{figure}
    \centering
    \includegraphics[scale = 0.42]{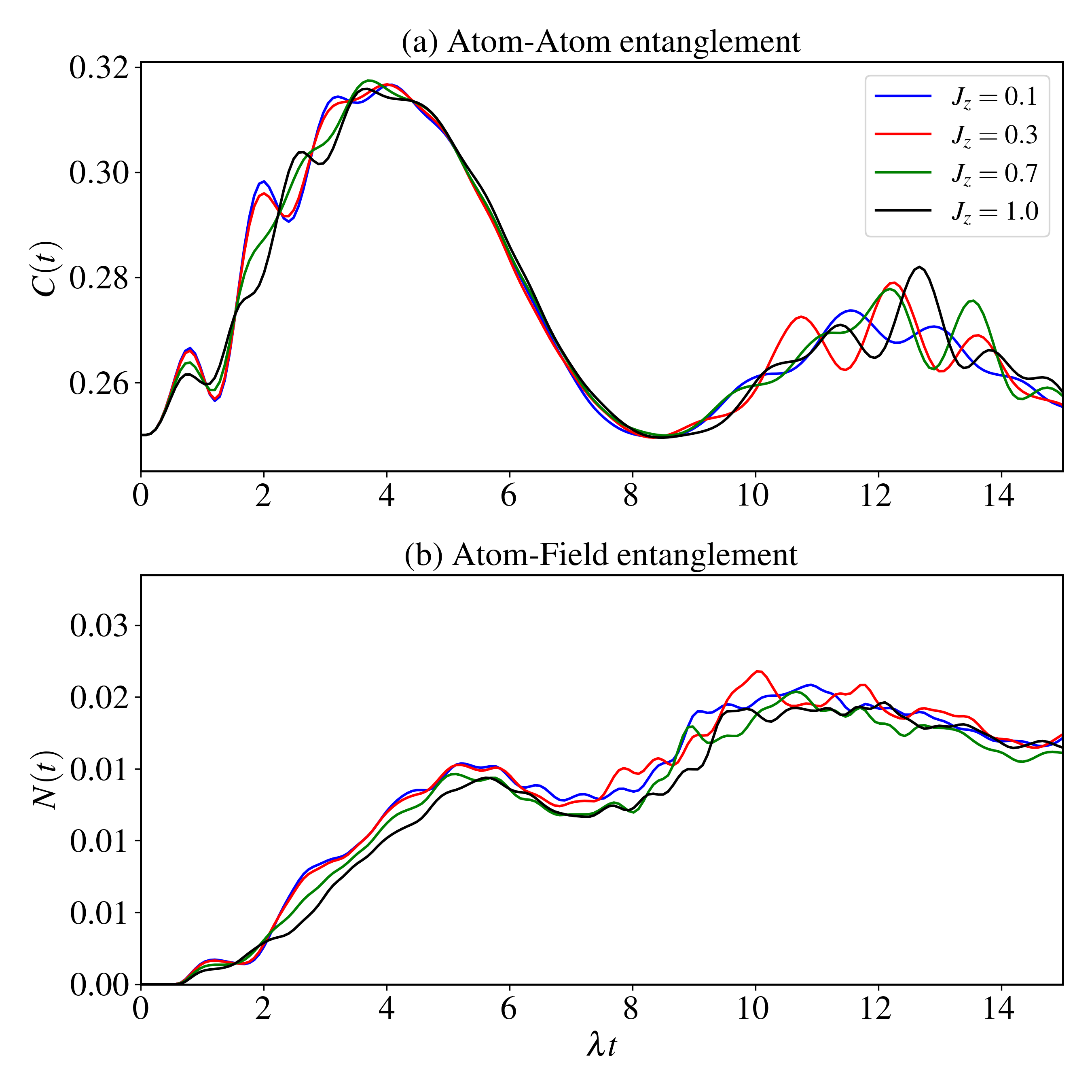}
    \caption{Effects of Ising-type interaction on entanglement dynamics with atoms in Werner state ( $\eta = 0.5$) and field is in SCTS. The parameters used in this plot are $\bar{n}_c = 5, \bar{n}_{th} = 1, \bar{n}_s = 1$ and  $J_{z} = 0.1, 0.3, 0.7, 1.0$. Blue curve $\Rightarrow J = 0.1$, red curve $\Rightarrow J = 0.3$, green curve $ \Rightarrow J = 0.7$ and black curve $\Rightarrow J = 1.0$. }
    \label{fig_10}
\end{figure}

\noindent{\textit{\textbf{Atoms in Werner state:}}}\\
The influence of the Ising-type interaction on the entanglement dynamics for atoms initially prepared in a Werner state is displayed in Fig.~\ref{fig_10}. Similar to the Bell-state case, both the atomic concurrence $C(t)$ and the atom--field negativity $N(t)$ remain largely unaffected by the inclusion of the Ising coupling. Although minor quantitative changes can be observed as $J_z$ increases, the overall dynamical features, such as oscillation patterns and amplitudes, remain essentially unchanged. This further confirms that, within the present model, the Ising-type interaction plays a secondary role in controlling entanglement dynamics compared to excitation-exchange interactions such as dipole--dipole coupling which is discussed below.
\section{Effects of dipole-dipole interaction}
The effects of the dipole-dipole interaction on the entanglement dynamics of atom-atom and atom-field subsystems are investigated in this section. The Hamiltonian including dipole-dipole interaction is  
\begin{equation}
\hat{H}_{\text{dd}} = \hat{H} + g_{d} \left(\hat{\sigma}^{\text{A}}_{+} \hat{\sigma}^{\text{B}}_{-} + \hat{\sigma}^{\text{A}}_{-} \hat{\sigma}^{\text{B}}_{+} \right),
\end{equation}  
where \(g_d\) is the dipole-dipole coupling strength. This term enables direct excitation exchange between the atoms, providing an additional coherent channel independent of the cavity field.

Dipole-dipole interactions are known to modify entanglement and revival phenomena in atom-field systems. For instance, in Ref.~\cite{PhysRevA.44.2135}, it was shown that two-atom systems with dipole-dipole coupling exhibit multiple revival sequences due to the interplay between single-atom Rabi oscillations and coherent atomic excitation exchange. Unlike the single-atom case with a Gaussian collapse envelope, the collapse and revival patterns become more complex, featuring partial and overlapping revivals. Evseev et al.~\cite{Evseev_2017} demonstrated that in non-resonant DJCMs, dipole-dipole interactions strongly influence concurrence oscillations. More recently, Mandal et al.~\cite{mandal2024entanglement} showed that in intensity-dependent DJCMs, dipole-dipole interactions can suppress entanglement sudden death (ESD), a feature also observed in standard DJCMs for sufficiently strong interactions.\\

\begin{figure}
    \centering
    \includegraphics[scale = 0.42]{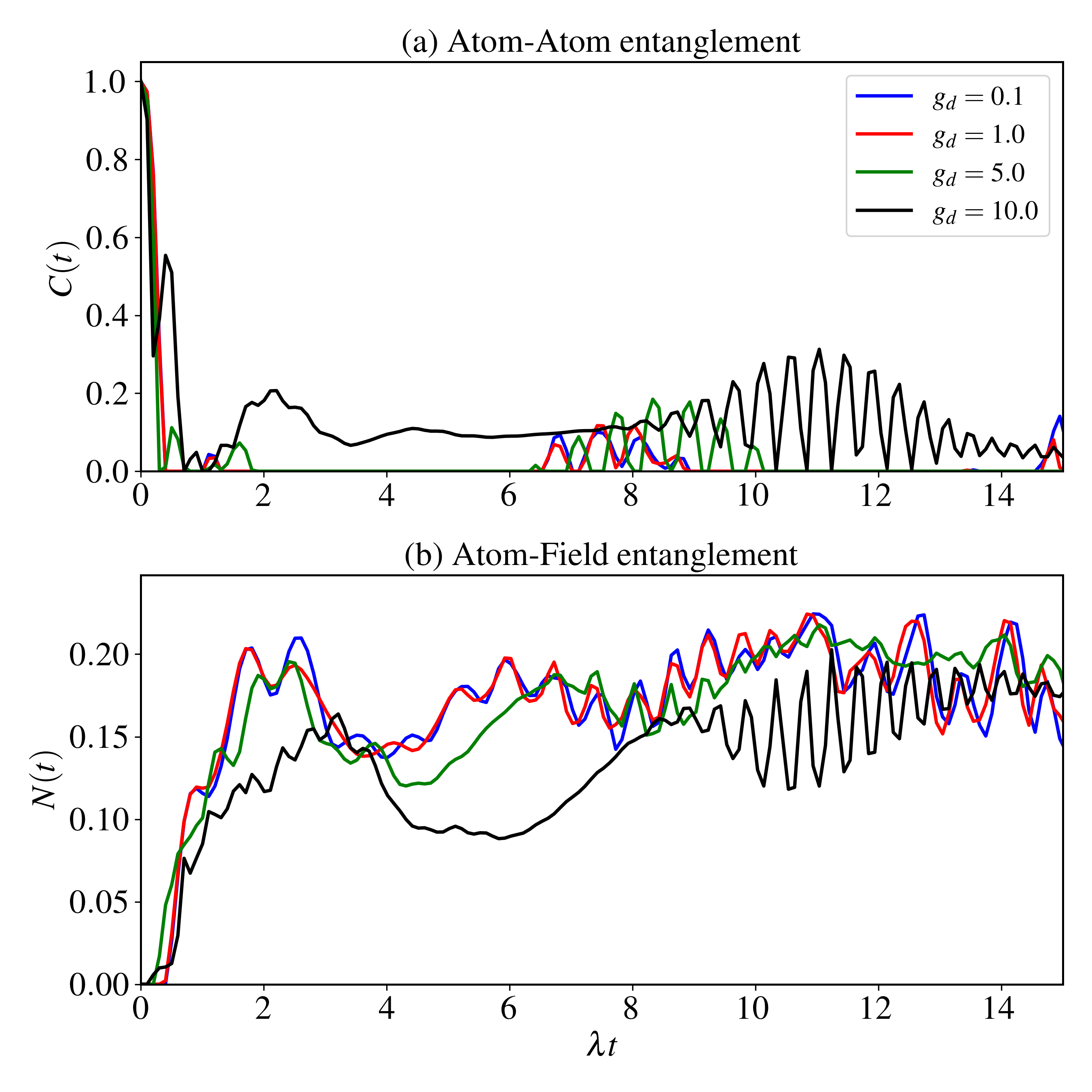}
    \caption{Effects of dipole-dipole interaction on entanglement dynamics with atoms in Bell state and radiation field is in SCTS. The parameters used in this plot are $\bar{n}_c = 5, \bar{n}_{th} = 1, \bar{n}_s = 1$ and  $g_d = 0.1, 1.0, 5.0, 10.0$. Blue curve $\Rightarrow g_d = 0.1$, red curve $\Rightarrow g_d = 1.0$, green curve $ \Rightarrow g_d = 5.0$ and black curve $\Rightarrow g_d = 10.0$. }
    \label{fig_11}
\end{figure}

\noindent{\textit{\textbf{Atoms in Bell state:}}} \\ 
Figure~\ref{fig_11} shows the atomic concurrence \(C(t)\) and atom-field negativity \(N(t)\) for atoms initially in a Bell state, with the cavity field in a single coherent-type state (SCTS). The curves correspond to different dipole-dipole couplings: \(g_d=0.1\) (blue), \(g_d=1\) (red), \(g_d=5\) (green), and \(g_d=10\) (black). For weak (\(g_d=0.1\)) and moderate (\(g_d=1\)) couplings, the dynamics of \(C(t)\) and \(N(t)\) closely follow the case without dipole-dipole interaction, indicating that atom-field interactions dominate and the direct atom-atom exchange has little effect. The concurrence exhibits standard oscillations with periodic revivals, while the negativity remains nearly in phase with the concurrence.

For strong (\(g_d=5\)) and very strong (\(g_d=10\)) couplings, significant changes emerge. In particular, for \(g_d=10\) (black curve), the amplitude of \(C(t)\) increases, and some ESD intervals are eliminated, as the strong dipole-dipole interaction provides an alternative coherent channel for entanglement, preserving atomic correlations. Conversely, \(N(t)\) decreases for larger \(g_d\), reflecting that the atom-atom subsystem increasingly forms a quasi-isolated entangled pair, thereby reducing atom-field correlations. The oscillation frequency of \(C(t)\) also increases with \(g_d\), consistent with faster coherent energy exchange between the atoms.\\

\begin{figure}
    \centering
    \includegraphics[scale = 0.42]{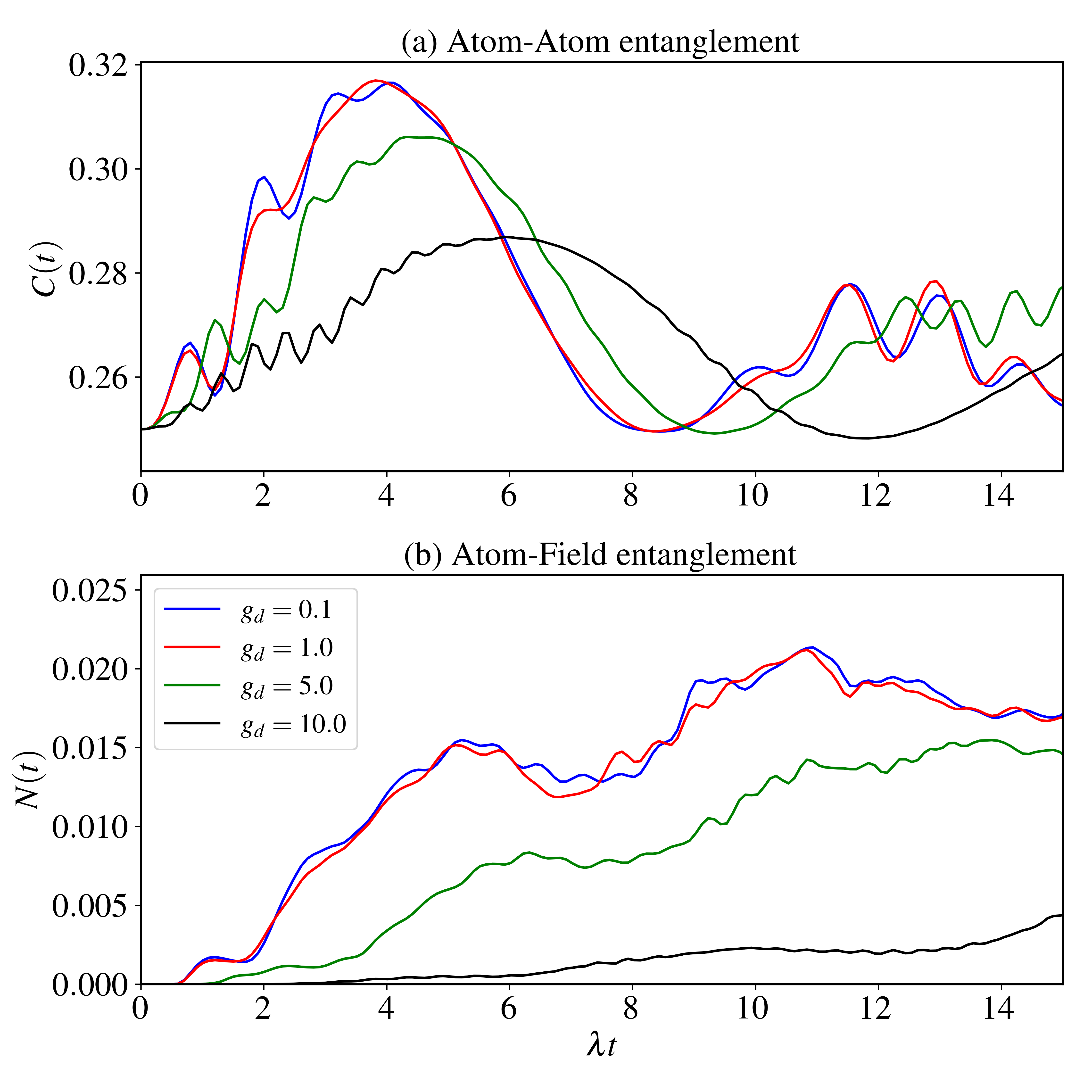}
    \caption{Effects of dipole interaction on entanglement dynamics with atoms in Werner state ($\eta = 0.5$) state and field is in SCTS. The parameters used in this plot are $\bar{n}_c = 5, \bar{n}_{th} = 1, \bar{n}_s = 1$; $\eta = 0.5$ and  $g_d = 0.1, 1.0, 5.0, 10.0$. Blue curve $\Rightarrow g_d = 0.1$, red curve $\Rightarrow g_d = 1.0$, green curve $ \Rightarrow g_d = 5.0$ and black curve $\Rightarrow g_d = 10.0$. }
    \label{fig_12}
\end{figure}
\noindent{\textit{\textbf{Atoms in Werner state:}}}\\  
Figure~\ref{fig_12} presents the dynamics for atoms initially in a Werner state. For weak and moderate couplings (\(g_d=0.1\) blue, \(g_d=1\) red), both \(C(t)\) and \(N(t)\) remain almost unchanged, demonstrating that the initial mixedness limits the effect of weak dipole-dipole coherence. For stronger couplings (\(g_d=5\) green, \(g_d=10\) black), both concurrence and negativity decrease noticeably. The reduction of \(C(t)\) indicates that strong interatomic interaction can partially redistribute correlations, while the decrease in \(N(t)\) signifies suppressed atom-field entanglement. Oscillation patterns also show reduced amplitude and faster cycles at high \(g_d\), illustrating enhanced coherent exchange between the atoms.

In summary, these results demonstrate that the dipole-dipole interaction acts as a tunable knob to control the entanglement structure. Weak interactions are largely negligible, while strong interactions enhance atomic entanglement, reduce atom-field correlations, modify oscillation frequencies, and can suppress ESDs in specific regimes.

\section{Effects of detuning}
In this section, we examine the influence of atom--field detuning on the entanglement dynamics of the atom--atom and atom--field subsystems. Detuning plays a central role in cavity and circuit QED systems, as it modifies the effective atom--field coupling strength and alters the rate of coherent energy exchange. In Refs.~\cite{Mandal_2024, mandal2024entanglement}, we previously investigated the impact of detuning on the concurrence $C(t)$ and negativity $N(t)$ in the double Jaynes--Cummings model (DJCM) and its intensity-dependent extension.

Including detuning, the effective Hamiltonian of the system can be written as
\begin{equation}
\hat{H}'_{\text{eff}} =
\Delta\, \hat{\sigma}_{-}^{\text{A}} \hat{\sigma}_{+}^{\text{A}}
+ \lambda \left(\hat{a}^{\dagger} \hat{\sigma}_{-}^{\text{A}} + \hat{a}\, \hat{\sigma}_{+}^{\text{A}}\right)
+ \Delta\, \hat{\sigma}_{-}^{\text{B}} \hat{\sigma}_{+}^{\text{B}}
+ \lambda \left(\hat{a}^{\dagger} \hat{\sigma}_{-}^{\text{B}} + \hat{a}\, \hat{\sigma}_{+}^{\text{B}}\right),
\label{detuningHamiltonian}
\end{equation}
where $\Delta = \omega - \nu$ denotes the atom--field detuning and $\lambda$ is the atom--field coupling strength. The detuning term introduces a mismatch between the atomic transition frequency and the cavity mode, effectively reducing resonant energy exchange while inducing phase-dependent dynamics.\\

\begin{figure}
    \centering
    \includegraphics[scale = 0.45]{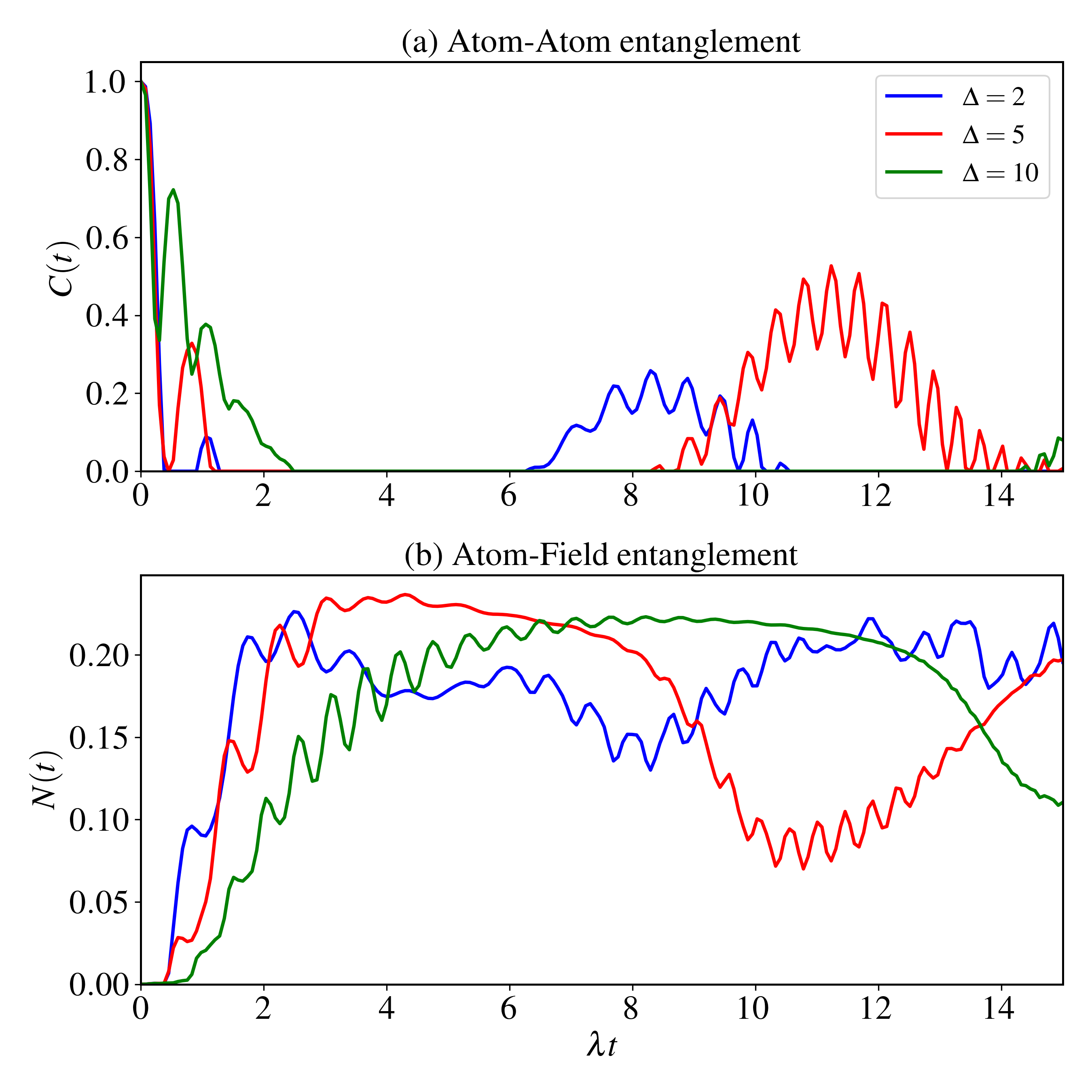}
    \caption{Effects of detuning on entanglement dynamics with atoms in Bell state and field is in SCTS. The parameters used in this plot are $\bar{n}_c = 5, \bar{n}_{th} = 1, \bar{n}_s = 1$ and  $\Delta = 2, 5, 10$. Blue curve $\Rightarrow \Delta = 2$, green curve $\Rightarrow \Delta = 5$, red curve $\Rightarrow \Delta = 10$.}
    \label{fig_13}
\end{figure}

\noindent{\textit{\textbf{Atoms in Bell state:}}}\\
The effects of detuning on the entanglement dynamics for atoms initially prepared in a Bell state are illustrated in Fig.~\ref{fig_13}. As seen in Fig.~\ref{fig_13}(a), the presence of detuning with $\Delta=2$ enhances the amplitude of the atomic concurrence $C(t)$ compared to the resonant case (cf. the blue curve in Fig.~\ref{fig_13}(a) and the black curve in Fig.~\ref{fig_1}(a)). This enhancement arises because moderate detuning suppresses rapid energy leakage into the field, thereby protecting atom--atom correlations.

When the detuning is increased to $\Delta=5$, the initial entanglement sudden death (ESD) is removed and a pronounced revival peak emerges in the dynamics. For even larger detuning ($\Delta=10$), the concurrence increases further and the early-time ESDs are completely eliminated, although ESDs may still occur at later times. This behavior indicates that detuning can act as a partial entanglement-preserving mechanism by effectively decoupling the atoms from the cavity field and slowing down decoherence pathways.

In contrast, the atom--field entanglement $N(t)$ exhibits an opposite trend, as shown in Fig.~\ref{fig_13}(b). Increasing detuning leads to longer initial ESD intervals and a substantial reduction in the early-time amplitude of $N(t)$, reflecting the weakened atom--field interaction. However, at later times, the amplitude of $N(t)$ increases for larger values of $\Delta$, suggesting a delayed but non-negligible buildup of atom--field correlations. These results clearly demonstrate the complementary nature of atom--atom and atom--field entanglement in the presence of detuning.\\

\begin{figure}
    \centering
    \includegraphics[scale = 0.45]{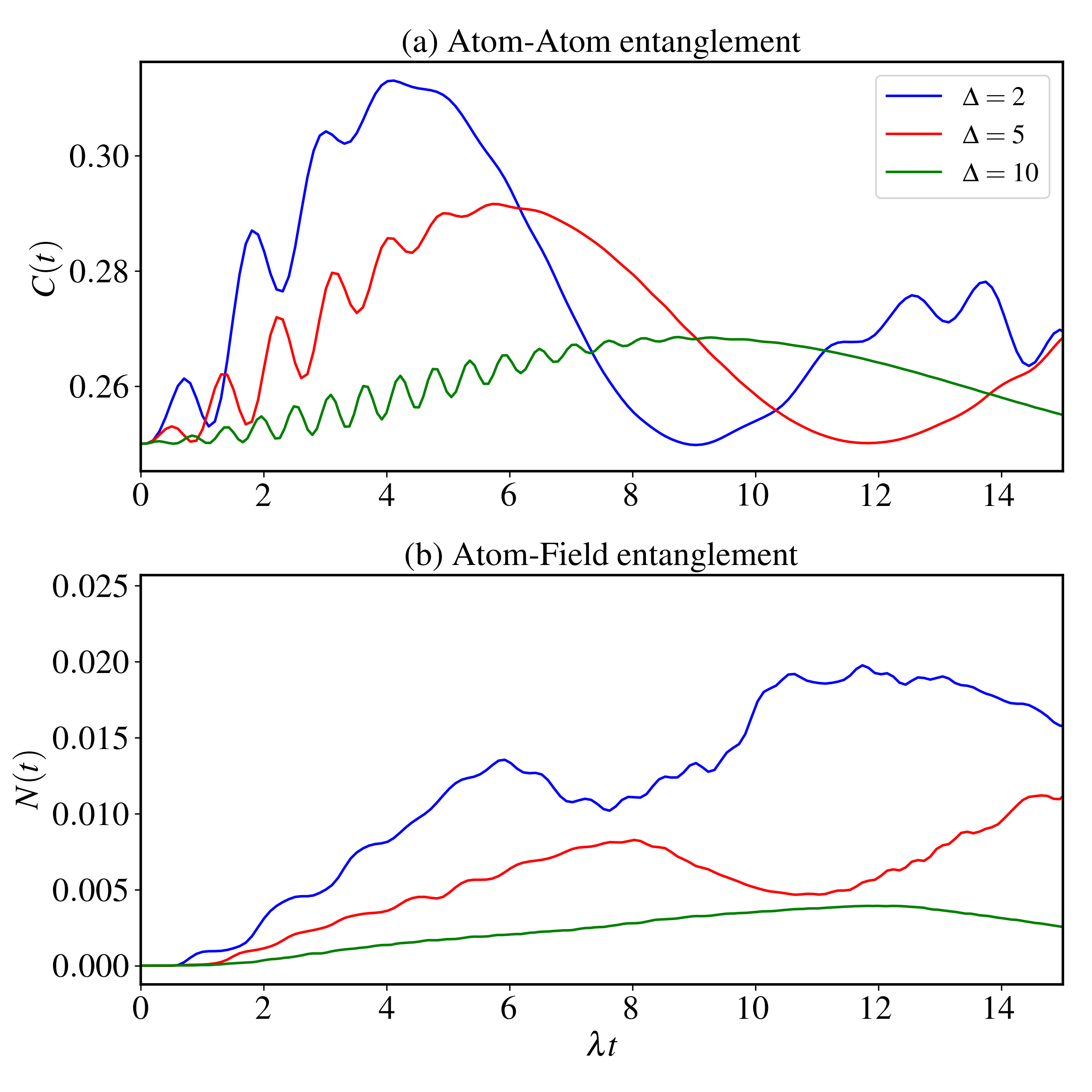}
    \caption{Effects of detuning on entanglement dynamics with atoms in Werner state ($\eta = 0.5$) and field is in SCTS. The parameters used in this plot are $\bar{n}_c = 5, \bar{n}_{th} = 1, \bar{n}_s = 1$ and  $\Delta = 2, 5, 10$. Blue curve $\Rightarrow \Delta = 2$, red curve $\Rightarrow \Delta = 5$, and black curve $\Rightarrow \Delta = 10$. }
    \label{fig_14}
\end{figure}

\noindent{\textit{\textbf{Atoms in Werner state:}}}\\
Figure~\ref{fig_14} shows the effect of detuning on entanglement dynamics for atoms initially prepared in a Werner state. In sharp contrast to the Bell-state case, increasing detuning leads to a significant suppression of the atomic concurrence $C(t)$. Specifically, the amplitude of $C(t)$ for $\Delta=5$ is approximately half of that observed for $\Delta=2$, indicating that detuning adversely affects atomic entanglement when the initial state is mixed. This opposite response can be attributed to the fact that, in a Werner state, atomic correlations are already weakened by mixedness, and detuning further inhibits the indirect entanglement mediation via the cavity field.

The atom--field entanglement $N(t)$ exhibits a similar qualitative behavior for both Bell and Werner initial states. As shown in Fig.~\ref{fig_14}(b), increasing detuning reduces the magnitude of $N(t)$, with higher values of $\Delta$ leading to stronger suppression. This reflects the general tendency of detuning to weaken atom--field coupling and thereby reduce atom--field entanglement, regardless of the purity of the initial atomic state.

Overall, these results reveal that detuning acts as a versatile control parameter whose effect on entanglement strongly depends on the initial atomic state. While detuning can protect and enhance atom--atom entanglement for atoms initially in a pure Bell state, it suppresses atomic entanglement for mixed Werner states and universally diminishes atom--field correlations.

\section{Effects of Kerr-nonlinearity}

In this section, we examine the influence of Kerr-nonlinearity on the entanglement dynamics of the atom--atom and atom--field subsystems. The Hamiltonian of the atom--cavity system in the presence of a Kerr medium is given by
\begin{equation}
    \hat{H}_{\text{Kerr}} = \hat{H} + \chi \hat{a}^{\dagger 2} \hat{a}^{2},
\end{equation}
where $\chi = k\omega$ denotes the strength of the Kerr-nonlinear interaction and $k$ is a non-negative dimensionless parameter. The presence of Kerr-nonlinearity introduces photon--photon interactions inside the cavity, thereby modifying the atom--field coupling dynamics and, consequently, the entanglement properties of the system.

The role of Kerr-nonlinearity in quantum optical systems has attracted considerable attention, both in early studies and more recent works
\cite{PhysRevA.45.6816, PhysRevA.45.5056, PhysRevA.44.4623, ahmed2009dynamics, sivakumar2004nonlinear, Mo_2022, PhysRevB.105.245310, baghshahi2014entanglement, zheng2017intrinsic, PhysRevA.93.023844}. 
In addition to affecting atom--field interaction dynamics, Kerr-nonlinearity has been shown to strongly influence phase-space distributions such as the $Q$-function and Wigner function \cite{PhysRevA.44.4623}. The interplay between Kerr-nonlinearity and entanglement dynamics has also been explored in several contexts, including thermal fields \cite{mojaveri2018thermal} and nonlinear Jaynes--Cummings models \cite{thabet2019dynamics, HOU2006727}.  

\begin{figure}[ht!]
    \centering
    \includegraphics[scale = 0.4]{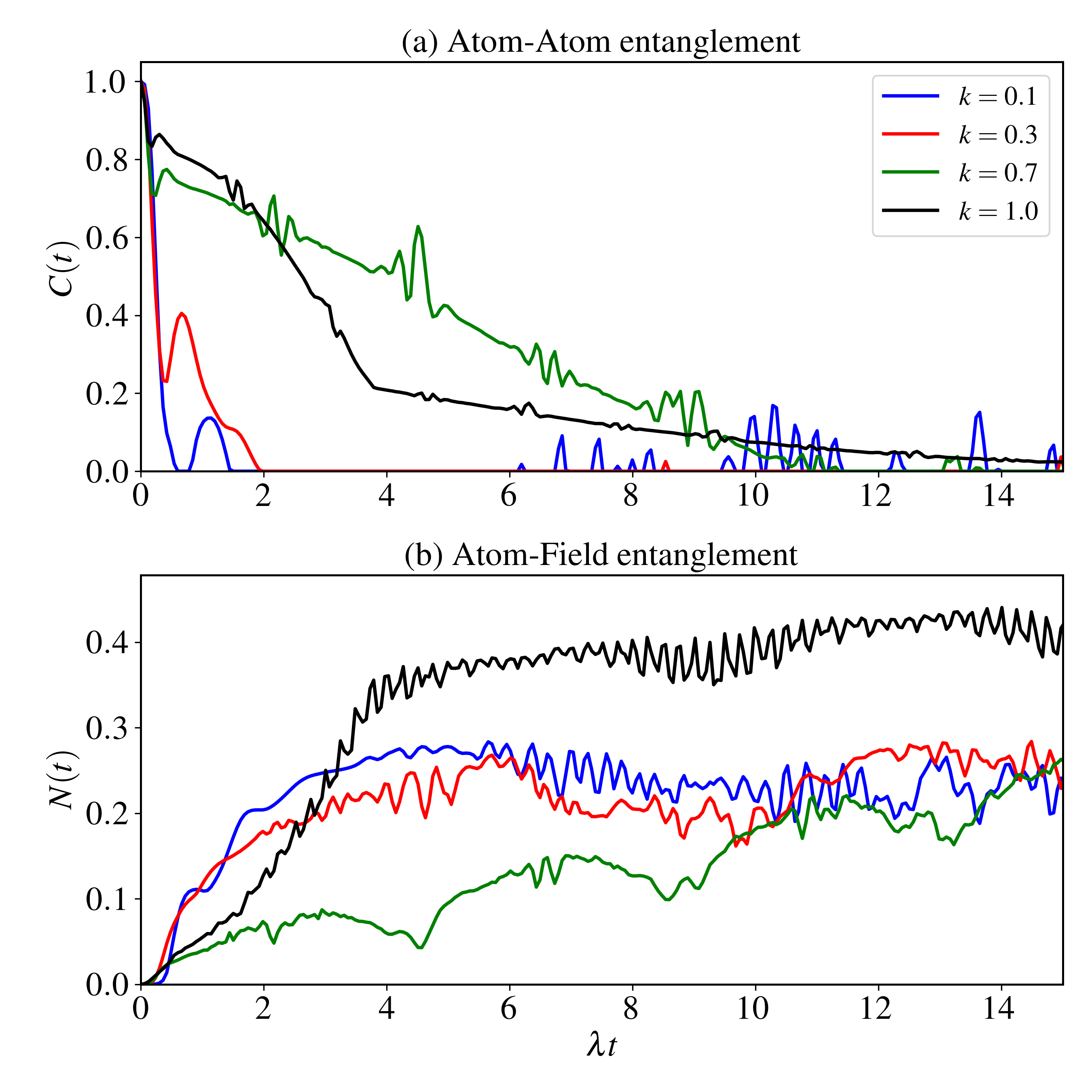}
    \caption{Effects of Kerr-nonlinearity on entanglement dynamics with atoms in Bell state and field is in SCTS. The parameters used in this plot are $\bar{n}_c = 5, \bar{n}_{th} = 1, \bar{n}_s = 1$ and  $k = 0.1, 0.3, 0.7, 1.0$. Blue curve $\Rightarrow k = 0.1$, red curve $\Rightarrow k = 0.3$, green curve $ \Rightarrow k = 0.7$ and black curve $\Rightarrow k = 1.0$. }
    \label{fig_15}
\end{figure}

\vspace{0.3cm}
\noindent\textbf{\textit{Atoms in Bell state:}}\\
The effects of Kerr-nonlinearity on the concurrence $C(t)$ and negativity $N(t)$ for atoms initially prepared in a Bell state are shown in Fig.~\ref{fig_15}. We observe that the inclusion of Kerr-nonlinearity significantly alters both the qualitative and quantitative features of the entanglement dynamics.

For weak nonlinearity ($k=0.1$), the duration of the initial entanglement sudden death (ESD) is slightly reduced, while the amplitudes of subsequent entanglement revivals are suppressed. As the Kerr strength is increased to $k=0.3$, the initial ESD region (up to $\lambda t \approx 2$) is completely removed; however, beyond this time interval, $C(t)$ rapidly decays and remains essentially zero. A further increase to $k=0.7$ leads to a dramatic enhancement of atom--atom entanglement: the amplitude of $C(t)$ increases substantially and all ESD events up to $\lambda t = 10$ are eliminated. This behavior persists and becomes more pronounced for $k=1.0$, indicating that strong Kerr-nonlinearity stabilizes atom--atom entanglement when the atoms are initially in a pure Bell state.

The atom--field entanglement $N(t)$ exhibits a complementary response to Kerr-nonlinearity. A comparison between the black curve in Fig.~\ref{fig_1}(b) and the blue curve in Fig.~\ref{fig_15}(b) reveals that a weak Kerr interaction ($k=0.1$) enhances $N(t)$. However, as $k$ is increased to $0.3$, the magnitude of $N(t)$ begins to decrease. Interestingly, for $k=0.7$, $N(t)$ is strongly suppressed at early times---coinciding with the enhancement of $C(t)$---but later regains strength as the dynamics evolves. This clearly demonstrates that Kerr-nonlinearity facilitates a redistribution of entanglement between the atom--atom and atom--field subsystems.

\vspace{0.3cm}

\begin{figure}[ht!]
    \centering
    \includegraphics[scale = 0.4]{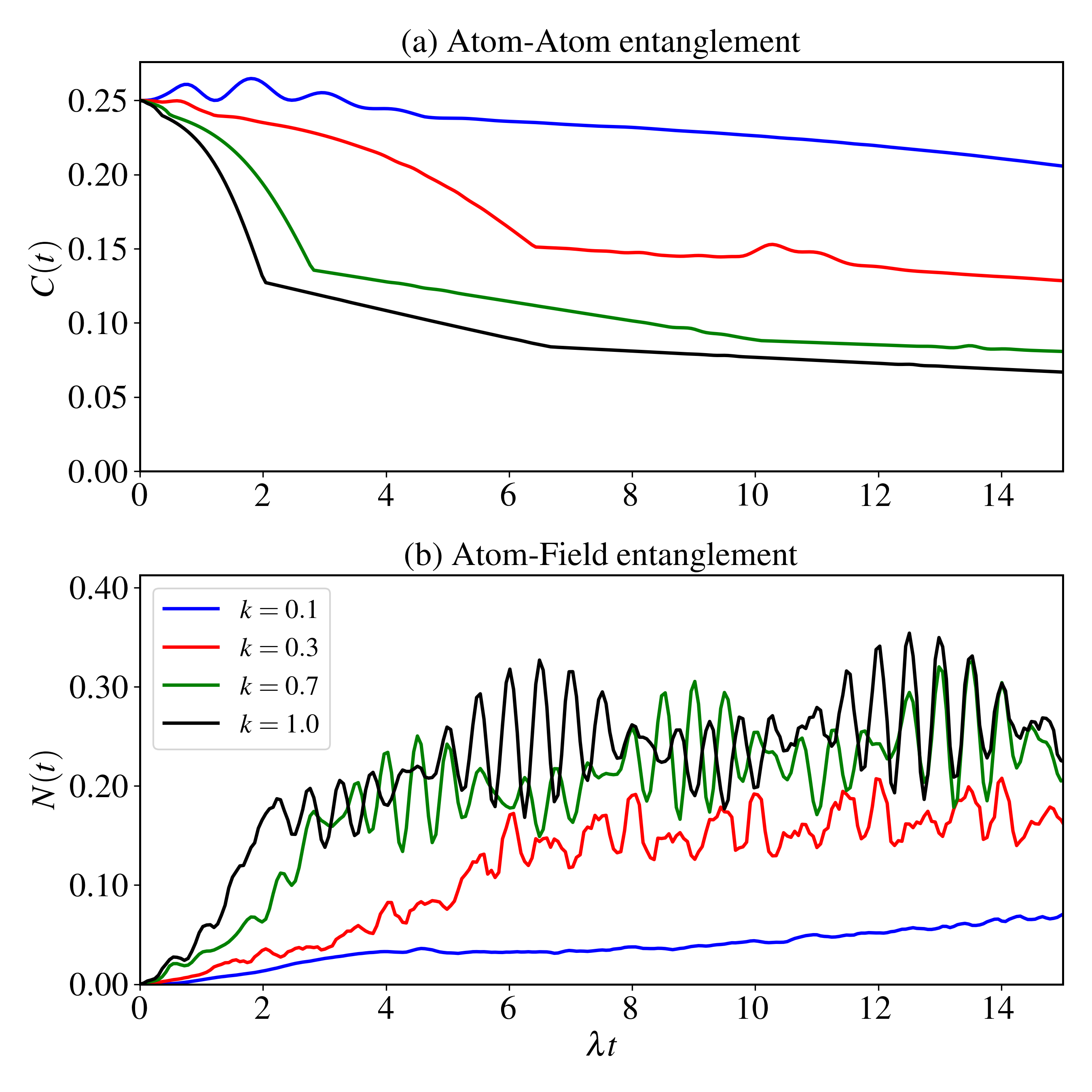}
    \caption{Effects of Kerr-nonlinearity on entanglement dynamics with atoms in Werner state and field is in SCTS. The parameters used in this plot are $\bar{n}_c = 5, \bar{n}_{th} = 1, \bar{n}_s = 1$, $\eta = 0.5$ and  $k = 0.1, 0.3, 0.7, 1.0$. Blue curve $\Rightarrow k = 0.1$, red curve $\Rightarrow k = 0.3$, green curve $ \Rightarrow k = 0.7$ and black curve $\Rightarrow k = 1.0$. }
    \label{fig_16}
\end{figure}

\noindent\textbf{\textit{Atoms in Werner state:}}\\
A qualitatively different behavior emerges when the atoms are initially prepared in a Werner state, as shown in Fig.~\ref{fig_16}. In contrast to the Bell-state case, the atom--atom entanglement $C(t)$ decreases monotonically with increasing Kerr strength $k$. For larger values of $k$, $C(t)$ exhibits a sharp initial decay, followed by a much slower reduction at later times. This indicates that strong Kerr-nonlinearity primarily affects atom--atom entanglement over short time scales, after which the dynamics stabilizes. Notably, despite this reduction in amplitude, no entanglement sudden death is observed for any value of $k$.

The atom--field entanglement $N(t)$, on the other hand, is significantly enhanced by Kerr-nonlinearity when the atoms are in a Werner state. Increasing $k$ leads to stronger oscillations in $N(t)$ and the formation of well-defined wave-packet structures, as evident from the black curve in Fig.~\ref{fig_16}(b). This enhancement reflects the increased role of photon--photon interactions in mediating atom--field correlations in the presence of mixed atomic states.

Overall, these results reveal that Kerr-nonlinearity acts as a powerful control parameter for directing entanglement flow within the system. Strong Kerr interaction favors robust atom--atom entanglement when the atoms are initially in a pure Bell state, whereas it enhances atom--field entanglement when the atoms are prepared in a mixed Werner state. This state-dependent response highlights the crucial interplay between initial atomic coherence and nonlinear cavity dynamics.

\section{Conclusion}

In this work, we have investigated the entanglement dynamics of atom--atom and atom--field subsystems within the Jaynes--Cummings model, with the radiation field initially prepared in a squeezed coherent thermal state. The atomic subsystem was taken to be either in a pure Bell state or a mixed Werner state, enabling a clear assessment of how atomic mixedness influences entanglement robustness and transfer.

In the absence of additional interactions, thermal photons were found to strongly suppress entanglement and significantly prolong entanglement sudden death (ESD), particularly for atom--atom entanglement when the atoms are initially in a Bell state. In contrast, squeezed photons enhance entanglement and partially counteract decoherence. When both thermal and squeezed photons are present, thermal noise dominates, leading to an overall reduction of entanglement in both subsystems.

The Wigner function analysis reveals a pronounced dependence on the initial atomic state. For Bell states, the field develops clear negativity, indicating the generation of nonclassical radiation. For Werner states, the Wigner function remains positive throughout the evolution, demonstrating that atomic mixedness suppresses field nonclassicality even in the presence of squeezing.

We further examined the influence of interaction-induced modifications to the Hamiltonian. Ising-type interaction was found to have only a marginal effect on entanglement dynamics for both subsystems. In contrast, dipole--dipole interaction enhances atom--atom entanglement and reduces ESD for Bell states, while suppressing entanglement for Werner states.

Detuning produces a strongly state-dependent response. For Bell states, increasing detuning mitigates ESD and enhances entanglement robustness, whereas for Werner states it significantly reduces entanglement amplitude. This complementary behavior highlights the interplay between atomic coherence and frequency mismatch.

Finally, Kerr nonlinearity was shown to profoundly reshape the entanglement landscape. For Bell states, strong Kerr nonlinearity enhances atom--atom entanglement and suppresses ESD, leading to long-lived correlations. For Werner states, it weakens atom--atom entanglement while enhancing atom--field entanglement, indicating a controllable transfer of entanglement between subsystems.

Overall, our results provide a comprehensive picture of how thermal noise, squeezing, nonlinearities, detuning, and interatomic interactions jointly govern entanglement dynamics in cavity QED systems. These findings offer useful guidance for controlling quantum correlations in realistic light--matter platforms where mixed states and environmental effects are unavoidable.

\section*{Acknowledgements}
The authors would like to thank Dr. Prabha Mandayam and Prof. Arul Lakshminarayan for their support and valuable discussions. We also would like to thank Dr. Chandrashekar R for his insightful suggestions and discussion.

\bibliographystyle{naturemag}
\bibliography{werner_jcm_scts.bib}

\end{document}